\documentclass[useAMS,usenatbib]{mn2e}
\usepackage{graphicx}
\usepackage{natbib}
\usepackage{threeparttable}

\newcommand{\Ras}{$R_{\rm in}~$} 
\newcommand{\TBL}{$T_{\rm BL}~$} 
\newcommand{\Rd}{$R_{\rm out}~$} 
\newcommand{\Rdp}{$R_{\rm out}$} 
\newcommand{\logg}{$\log g$}
\newcommand{\CDC}{\mbox{$^{12}$C/$^{13}$C}}

\newcommand{\Vt}{$V_{\rm t}$}

\newcommand{\Thp}{\mbox{$T_{\rm eff}^*$}}
\newcommand{\Tef}{\mbox{$T_{\rm eff}$}}

\newcommand{\Mdot}{\mbox{\,M$_\odot$~yr$^{-1}$}}

\newcommand{\Rsun}{\mbox{\,R$_\odot$}}
\newcommand{\vunit}{\mbox{\,km\,s$^{-1}$}}
\newcommand{\mic}{\mbox{$\,\mu$m~}}
\newcommand{\micp}{\mbox{$\,\mu$m}}
\newcommand{\pion}[2]{{#1}\,{\sc {#2}}}

\newcommand{\ltsimeq}{\raisebox{-0.6ex}{$\,\stackrel
        {\raisebox{-.2ex}{$\textstyle <$}}{\sim}\,$}}
\newcommand{\gtsimeq}{\raisebox{-0.6ex}{$\,\stackrel
        {\raisebox{-.2ex}{$\textstyle >$}}{\sim}\,$}}

\newcommand{\RS}{RS~Oph}

\begin{document}

 \title[Modelling the secondary in RS~Oph]
{Modelling the spectral energy distribution of the red giant
in \RS iuchi: evidence for irradiation}

\author[Ya. V. Pavlenko et al.]
{Ya. V. Pavlenko$^{1,2}$\thanks{E-mail:yp@mao.kiev.ua}, 
B. Kaminsky$^1$,
M. T. Rushton$^{3,4}$,
A. Evans$^5$,
C. E. Woodward$^6$, \newauthor
L. A. Helton$^6$\thanks{Present Address: SOFIA Science Center, USRA, NASA Ames
Research Center, M.S. 232-12, Moffett Field, CA 94035, USA},
T. J. O'Brien$^7$,
D. Jones$^{7,8}$,
V. Elkin$^3$\\ \\
\mbox{} 
$^{1}$Main Astronomical Observatory, Academy of 
Sciences of the Ukraine, Golosiiv Woods, Kyiv-127, 03680 Ukraine \\
$^2$ Center for Astrophysics Research, University of Hertfordshire,
College Lane, Hatfield, Hertfordshire AL10 9AB, UK \\
$^{3}$Jeremiah Horrocks Institute for Astrophysics and Supercomputing,
University of Central Lancashire, Preston, \\Lancashire PR1 2HE, UK\\
$^{4}$Astronomical Observatory of the Romanian Academy, Str. Cutitul de Argint 5, Bucharest, 040557, Romania\\
$^{5}$Astrophysics Group, Keele University, Keele, Staffordshire, ST5 5BG, UK \\
$^{6}$Minnesota Institute for Astrophysics, School of Physics and Astronomy,
116 Church Street, S. E., University of Minnesota,\\ Minneapolis, Minnesota 55455, USA\\
$^7$Instituto de Astrof\'isica de Canarias, E-38205 La Laguna, Tenerife, Spain\\
$^8$Departamento de Astrof\'isica, Universidad de La Laguna, E-38206 La Laguna, Tenerife, Spain\\
}

\date{}
\pagerange{\pageref{firstpage}--\pageref{lastpage}} \pubyear{2002}
\maketitle
\label{firstpage}

\begin{abstract}
We present an analysis of optical and infrared spectra of the
recurrent nova RS Oph obtained during between 2006 and 2009. 
The best fit to the optical spectrum for 2006 September 28  
gives \Tef = $3900$~K for $\log{g} = 2.0$, while for $\log{g} = 0.0$ 
we find \Tef = $4700$~K,  and a comparison with template stellar spectra 
provides \Tef $\sim$ 4500 K. The observed spectral energy 
distribution (SED), and the intensities of the emission 
lines, vary on short ($\ltsimeq 1$~day) time-scales, due
 to disc variability.
We invoke a simple 
one-component model for the accretion disc, and a model with 
a hot boundary layer, with 
high ($\sim 3.9 \times 10^{-6}$ \Mdot) and low ($\sim 2 \times 10^{-8}$ 
\Mdot) accretion rates, respectively. Fits to the accretion disc-extracted 
infrared spectrum (2008 July 15) yield effective temperatures for the 
red giant of $\Tef = 3800 \pm 100$~K ($\log{g} = 2.0$) and $\Tef = 
3700 \pm 100$~K ($\log{g} = 0.0$). Furthermore, 
using a more sophisticated approach, we reproduced the optical and infrared SEDs of 
the red giant in the RS Oph system with a two-component model atmosphere, in which 90\% of the surface 
has \Tef = 3600 K and 10\% has \Tef = 5000~K. 
Such structure could be due to irradiation of the red giant by the white dwarf.
\end{abstract}

\begin{keywords}
stars: individual (RS Oph) ---
binaries: symbiotic ---
novae, cataclysmic variables ---
stars: effective temperatures ---
circumstellar matter
\end{keywords}

\section{INTRODUCTION}
 \label{intro}
RS~Ophiuchi is a recurrent nova (RN), an interacting binary system in 
which multiple nova outbursts have been observed; it has undergone 
eight known or suspected outbursts since 1898 \citep{anupama08}, the 
most recent peaked on 2006 February $12.94\pm0.04$ UT \citep{Hounsell10}.

\RS\ consists of a white dwarf (WD) primary close to the Chandrasekhar 
limit, accreting material from a red giant \citep[RG; spectral type 
M2IIIpe+, see {\it SIMBAD}\footnote{http://simbad.u-strasbg.fr/Simbad} 
and][]{fekel00} secondary star. The RN eruption is explained in terms of 
a thermonuclear runaway on the surface of the 
WD \citep[e.g.,][]{starrfield72, starrfield08} following 
the accretion of sufficient mass by the WD to trigger ignition. In quiescence, 
the optical spectrum is dominated mainly by the RG. 
The variable observed hot 
component in \RS\ spectrum can be explained by a white dwarf+accretion 
disc (AD) embedded in an envelope of wind from the M giant secondary. 
The observed variations of its hot component are a result of (a) fluctuations in 
the mass accretion rate; (b) changes in the column density of the 
absorbing wind envelope, which is optically thick 
(\cite{dobrzycka96, anupama99}).

Unlike the classical novae, the masses of the WD in RNe may be increasing over 
time. If the WD is a CO WD, then the mass of the 
WD in \RS\ may eventually reach
the Chandrasekhar limit and explode as a Type Ia supernova 
\citep{IbenTut1984, wood}. 
Whether or not RS Oph is gaining mass remains a matter of considerable
debate \citep[e.g.,][]{ness09}.

\cite{wallerstein08} determined elemental abundances for the donor stars of 
RS Oph and the recurrent nova T~CrB
(which also has a red giant secondary; see \citealt{anupama08}).
They analysed the $6500-8800$\,\AA\ spectral range, using  
\pion{Fe}{i}, \pion{Ni}{i}, \pion{Si}{i} and \pion{Ti}{i} lines to establish 
effective temperatures of $4100-4400$\,K (RS~Oph) and 3600\,K (T~CrB). 
From analysis of the equivalent widths of atomic 
lines they determined a near-solar metallicity for both stars, but with large
(factor $\sim2$) uncertainties. Similarly, \cite{pavlenko08_rsoph} analysed infrared
(IR) spectra of \RS\ and determined $\Tef = 4100\pm100$~K for the secondary.
They also found that the metallicity of the secondary is solar, but with a deficit
of carbon and an overabundance of nitrogen with respect to solar (see also \citealt{scott}). 

A nitrogen overabundance and a carbon underabundance is known in a variety of
the normal RGs, with a range of metallicities \citep[][ and references
therein]{smith05}, and is due to the conversion of C to N in the CN-cycle of
hydrogen burning, and the subsequent dredge-up of the products on the Red Giant
Branch \citep[e.g.,][]{smithlambert85,smithlambert86,smiljanic09}. Therefore,
the material accreting onto the WD from the M giant secondary in 
the \RS\ system likely is N-enhanced \citep[see Fig.~14 of][]{ness09}. 

\cite{pavlenko10} carried out an analysis of a high resolution IR spectrum of 
the RG component of \RS\ and found $\CDC =16 \pm 3$. This value is consistent
with those typically observed in the atmospheres of RG stars in open 
clusters after first dredge up \citep{smiljanic09}. Likely,
the RG in \RS\ has undergone some further mixing, 
 suppressing the \CDC\ ratio still further.

Dust is present in the \RS\ system \citep{evans07, vanloon08, woodward08},
emitting primarily at $\lambda \gtsimeq 10$\micp. \cite{rushton10} 
have modelled the spectral energy distribution (SED) of \RS\ in the near-IR
(1--5\mic) and found that \Tef\ is essentially constant over the period 2006
August -- 2008 July. \cite{rushton10} have shown that there also is evidence for
dust emission in the near-IR at  3\mic\ -- 5\micp. Observations at longer
wavelengths clearly show that this emission by the dust envelope is
variable (Rushton et al., to be submitted). 

In this paper we do not consider the effect of the dust on the
optical spectra: \cite{pavlenko08_rsoph} and \cite{rushton10} 
have shown that even the near-IR spectrum of \RS\ is not affected
by emission from the  hot ($\sim500-600$~K) dust known
to reside in the \RS\ environment, and which is
likely a permanent feature of the system.

\section{OBSERVATIONS AND REDUCTION}

\subsection{Optical spectra\label{_ostv}}

An optical spectrum of \RS\ was obtained with the Boller \& Chivens (B\&C)
Spectrograph\footnote{http://james.as.arizona.edu/$\sim$psmith/90inch/90finst.html} 
at the Steward Observatory 2.3~m Bok Telescope on Kitt Peak, Arizona over
multiple epochs. Details of the observations are presented in Table~\ref{tab:bokobslog}.
 Also included in Table~\ref{tab:bokobslog} is the orbital phase $\phi$,
defined such that $\phi\simeq0.5\mbox{~and~}\sim1.0$ at quadrature.
 The ephemeris for the orbit is from \cite{fekel00},
essentially the same phases are obtained using the parameters in \cite{Brandi2009}.

All observations were obtained with a 400 line/mm 4889\AA\ first order 
grating, using a grating tilt of 6.5$^{\circ}$, a 1.5$^{\prime\prime}$ wide longslit, and 
appropriate order blocking filters. The scheme yielded a spectral
resolution at the detector of $\sim5$~\AA.

{\begin{table}
\caption{Bok (+B\&C Spectrograph) Observing Log
\label{tab:bokobslog}}
\centering
\begin{tabular}{ccccc}
\hline\hline

        & Average& & Total & Orbital \\
UT Date & Airmass& $N_{\rm combine}$ & Integration & Phase \\
        &        &  & sec & $\phi$ \\
\hline

2006 Sep 28.11 & 1.51& 24  & 1440 & 0.52 \\
2007 May 09.33 & 1.71& 14  & 840  & 0.98\\
2007 May 10.31 & 1.74& 38  & 2280 & 0.97\\
2008 Jun 25.19 & 1.84& 7   & 210  & 0.84 \\
2009 May 07.28 & 2.18& 10  & 225  & 0.50\\ 
2009 May 08.28 & 2.20& 10  & 450  & 0.48 \\

\hline\hline

\end{tabular}
\end{table}
}

The optical data were reduced in IRAF\footnote{IRAF is 
distributed by the National Optical Astronomy Observatory, which is operated by
the Association of Universities for Research in Astronomy, Inc., under
cooperative agreement with the National Science Foundation.} using standard  
two-dimensional optical spectroscopic data reductions techniques
\citep{Howell1992}. For each set of observations, multiple B\&C spectra were
obtained and combined with median filtering to enable removal of cosmic rays
prior to extraction. Wavelength calibration was performed using HeArNe
calibration lamps with exposures at the position of the target. Corrections for
pixel-to-pixel response variations (flatfield) were accomplished using a high
intensity quartz-halogen continuum lamp in the instrument. Spectra of
spectro\-photometric standard stars were obtained at airmasses similar to that
of \RS\ to correct for instrumental response and provide flux calibration. 

\begin{figure}
  
   \centering\includegraphics[width=80mm]{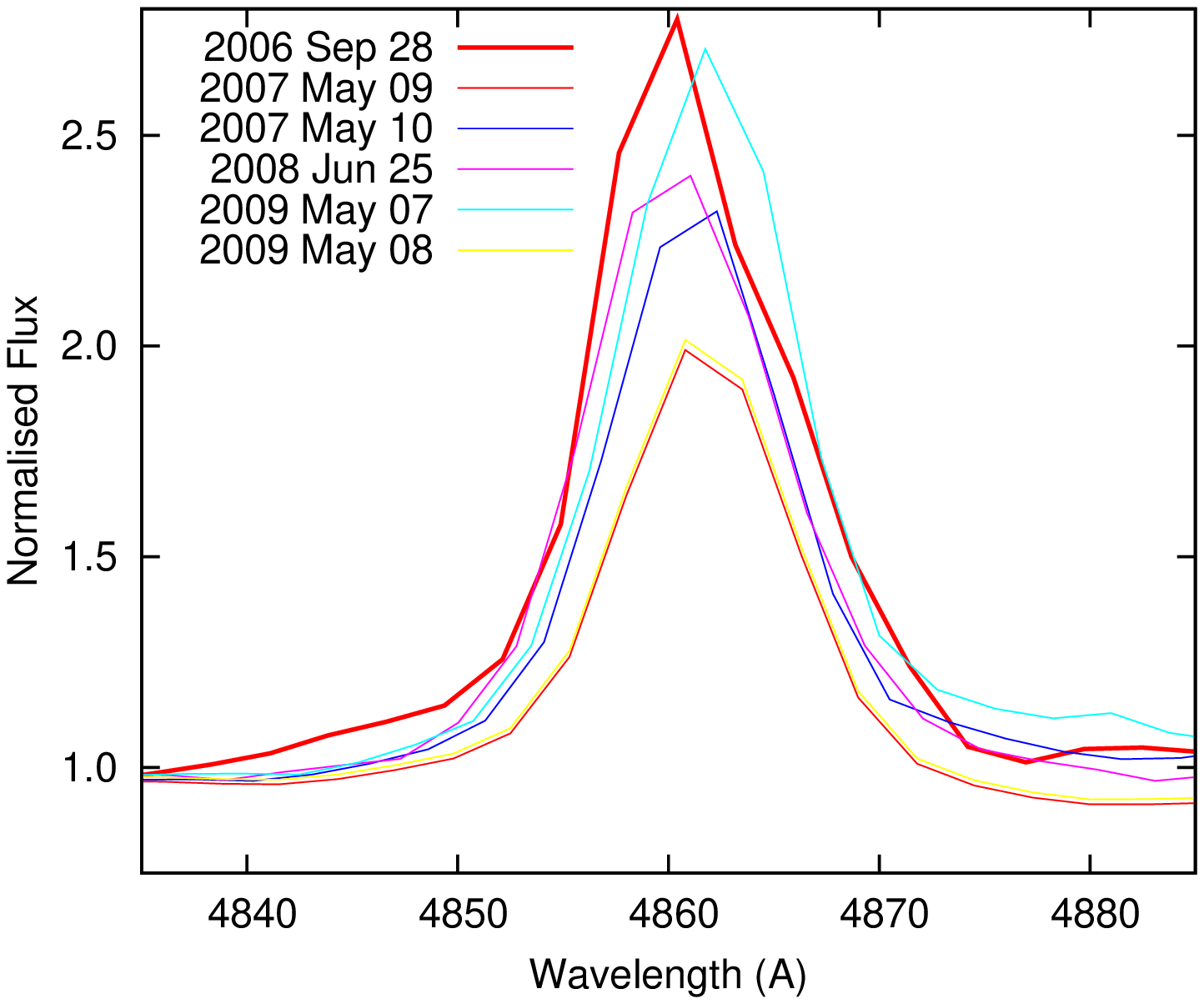}
   \centering\includegraphics[width=80mm]{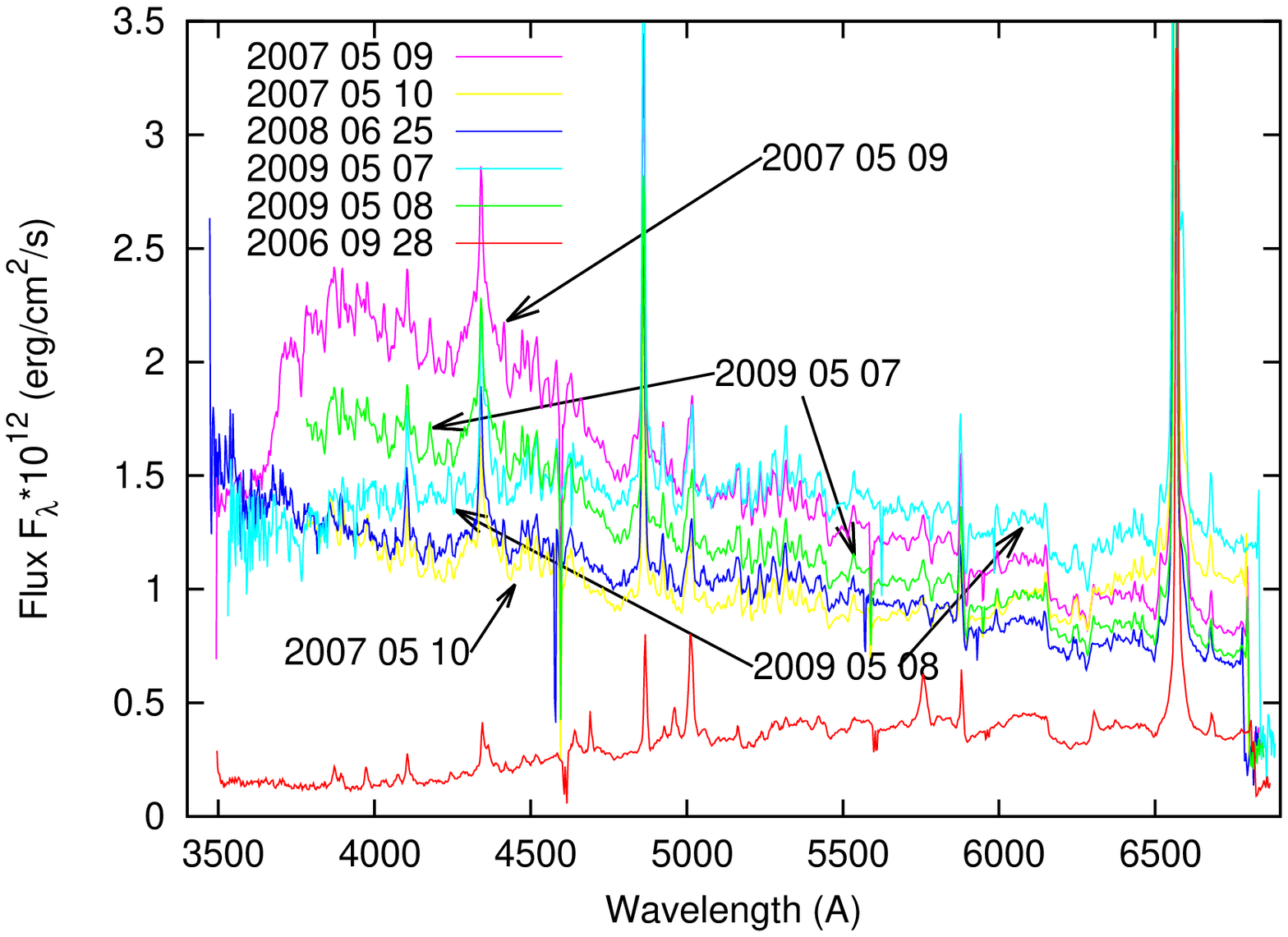}
      \caption{\label{_seds} 
      Top: Profiles of $H_{\beta}$ on different 
      dates. Fluxes are normalised to the same value at 4835\AA. 
            Bottom: comparison of the observed 
      fluxes in the 2006 September optical spectrum of RS Oph with the observed
      SEDs on the other dates. 
These and other figures are available in colour on 
ftp://ftp.mao.kiev.ua/pub/yp/Papers/2015rsoph}  
   \end{figure}

The low resolution Bok data provide some evidence for
variability of the spectra. In the  top panel of Fig.~\ref{_seds} we show 
 six
 spectra across $H_\beta$ line obtained on different UT dates. 
The spectra shown here are normalised to obtain similar pseudo\-continuum 
fluxes around the $H_{\beta}$ $\lambda=486.1$~nm line, seen in emission.
Despite the comparatively low resolution ($R = 1000$)
we see noticeable changes in the H$_\beta$ emission even on a 1-day time scale:
there are significant changes between 2007 May 9 and 10, and 2009 May 7 and 8 
(see upper panel of Fig.~\ref{_seds}). 
These variations in H$_\beta$ emission are greater
than the uncertainties in flux calibration, and we
consider them real.
 Furthermore, we note the similarity
      of $H_{\beta}$ on days 2007 May 9 and 2009 May 8, despite the difference of the
      observed SEDs on the same dates (see bottom panel of Fig.~\ref{_seds}).
      Indeed, we see a relatively stronger $H_{\beta}$ 
      when the disc is in a low state on day 2006 September 26
      (see Section~\ref{laterdates}).
In the bottom panel of Fig. ~\ref{_seds} we show the observed 
SEDs for the same six dates. Again, the
observed SEDs, as well as the $H_\beta$ line,
change on a time-scale $\sim$~ 1 day, the data for 2007 May 9 and May 10, and for
2009 May 7 and 8, are identified by arrows in the Figure.
 Optical flickering of RS Oph and SED variability are also reported by 
\cite{Worters07, voloshina, kundra} and \cite{zamanov}.

\subsection{IR spectrum\label{_irs}}

IR spectroscopy of \RS\  on various dates \citep[see Table 1 in][]{rushton10}
was obtained at the United Kingdom
Infrared Telescope (UKIRT), with the UKIRT Imager 
Spectrometer \citep[UIST;][]{ramsay}. The 
observations were obtained in stare-nod-along-slit mode, with
a 2-pixel-wide slit. We obtained data in the  different band grisms,
giving a spectral coverage of 0.86 -- 5.31\mic and resolution 
$ R~ \sim 600-2000$. We obtained spectra of HR 6493
(F3 V) immediately before we obtained spectra of \RS,
for calibration purposes. Detailed information about 
the observational procedure and reduction process is
given by \cite{rushton10} (see the Table 1 therein). For the present paper
we have used the spectrum obtained on 2008 July 25, being the nearest date to the optical 
spectrum; see Section~\ref{__fitsir}.

\section{The optical spectrum}

\subsection{The spectrum of \RS\ in 2006 September} \label{template}
   
\begin{figure}
   \centering
   \includegraphics[width=80mm]{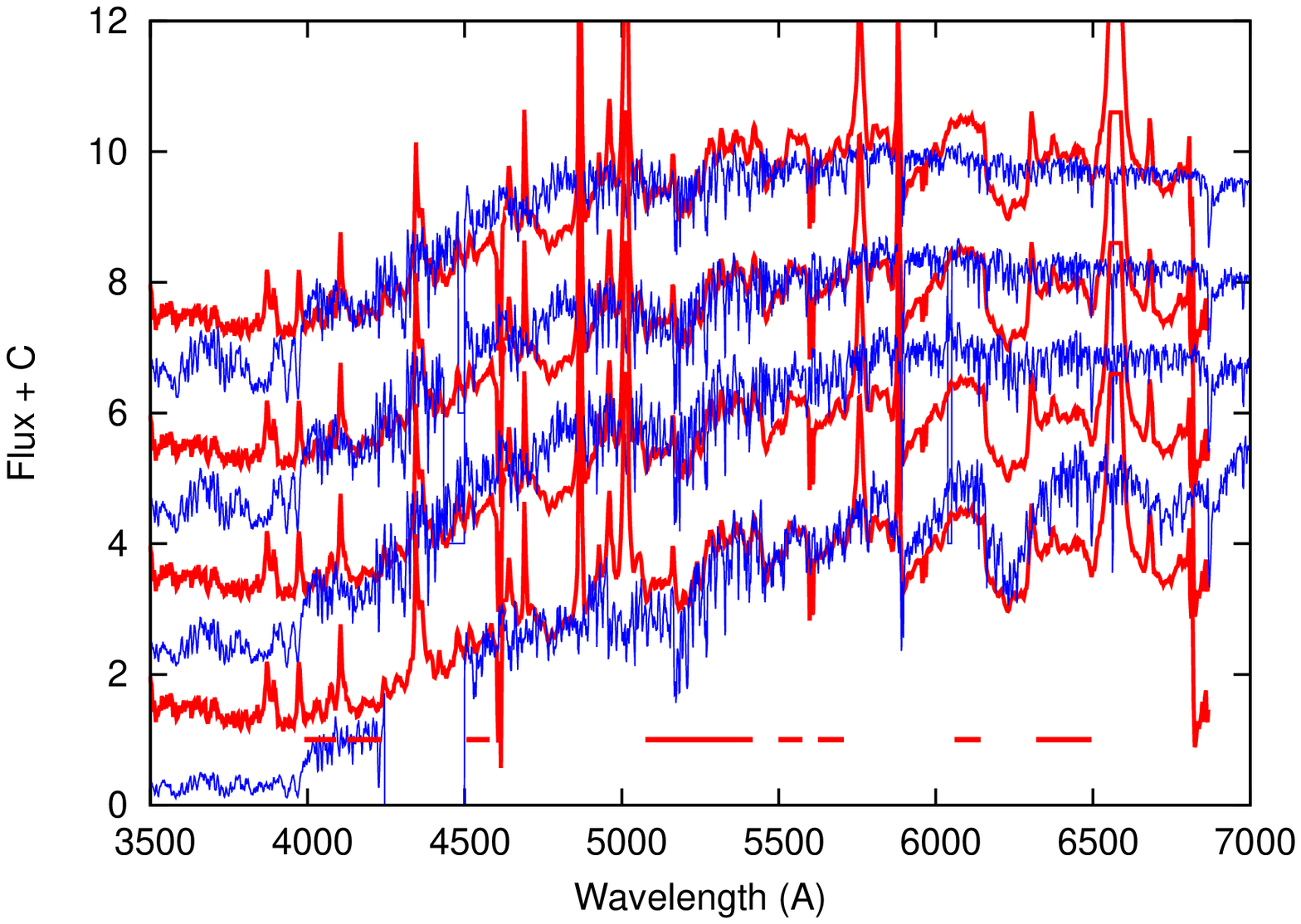}
   \includegraphics[width=80mm]{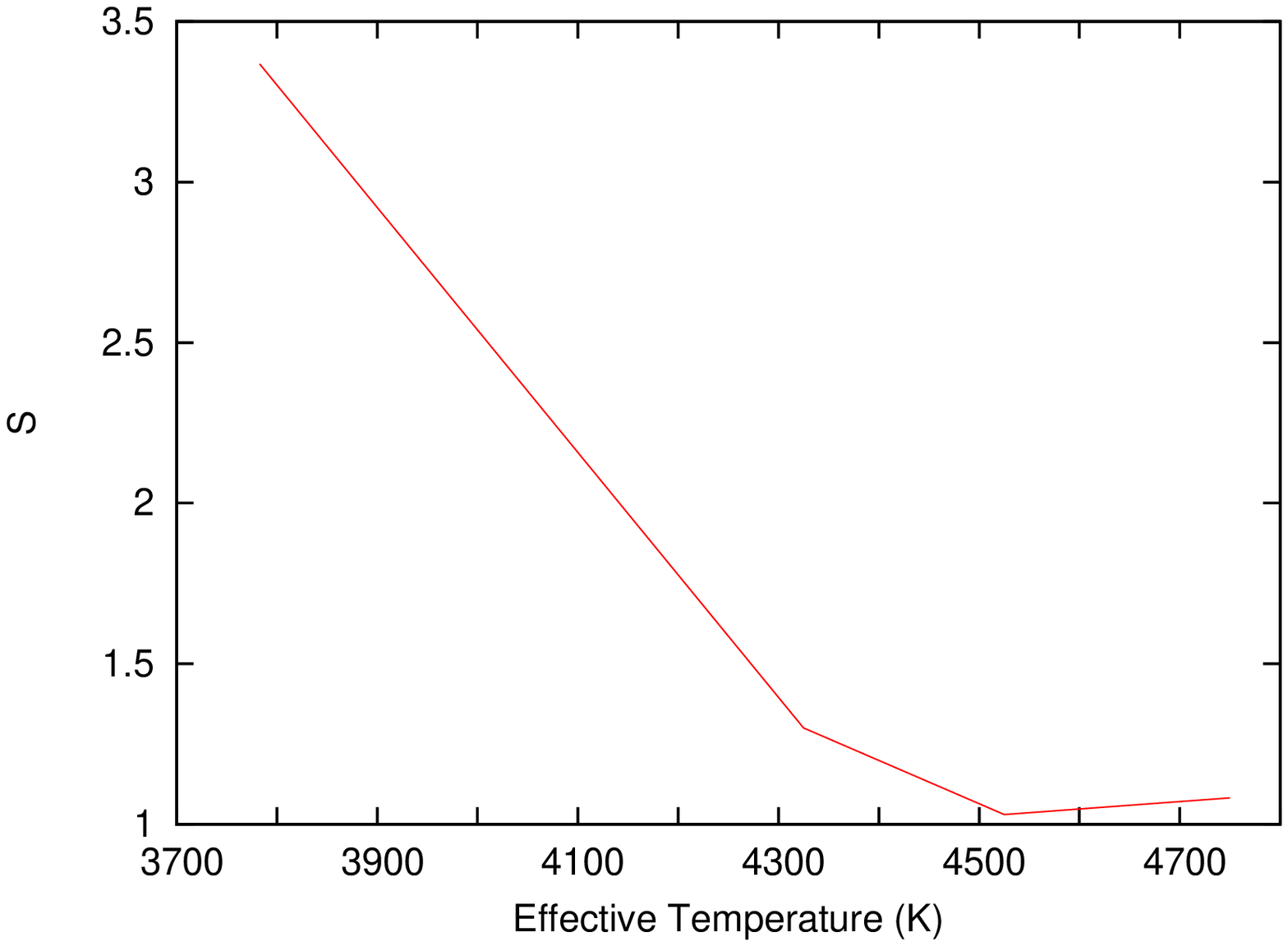}
\caption{\label{_templ}  Left: comparison of the observed 
fluxes in the 2006 Sep optical spectrum of RS Oph to the observed
SEDs of 4 template stars taken from the STELIB database 
(http://webast.ast.obs-mip.fr/stelib), from the top:
HD72324: \Tef/\logg/[Fe/H] = 4750/1.9/0, HD72184: 4525/2.05/-0.05, HD54719: 4325/2.17/0.02,
HD146051: 3783/0.0/0.  Right: result of the comparison of observed fluxes 
of \RS\ in Sept. 2006 with SEDs of the same template stars using $\chi^2$
criteria, see section \ref{cmod}. Spectral ranges of the continuum fluxes comparison are marked by
solid lines at Flux =1.0 on the left panel.}
\end{figure}

The earliest spectrum (2006 September) was obtained during the post-outburst
minimum, and before the resumption of optical flickering when the accretion disc (AD), which is
destroyed by the nova eruption, had not been fully re-established
\citep{Worters07}.  

We compare the observed SEDs in 2006 with similar 
data available for ``normal'' RGs.  Our rationale for this is that,
for our purposes, the definition of the SED relates more to the  continuum
fluxes formed at the level of the photosphere: molecular bands and atomic lines
form above the photosphere, where their formation may be affected by processes
such as irradiation, chromospheric-like features, etc. Our aim at this stage is
to determine effective temperatures.

We found the shape of the SED of \RS\ soon after the 2006 eruption seems
very similar to those of known RGs in the STELIB
database\footnote{http://webast.ast.obs-mip.fr/stelib}. A direct comparison of
the \RS\ SED from 2006 September with observed fluxes of 
the K-giants HD54719, HD72184 and HD72324 and the M-giant HD146051 
is shown in Fig.~\ref{_templ}.

The atmospheric parameters for the template stars were obtained from
\cite{massarotti08} (HD54719) and \cite{luck95} (HD72184, HD72324) 
and \cite{_koleva12} (HD146051). The 
effective temperatures are $\Tef = 4325$~K (HD54719), $\Tef = 4525$~K (HD72184) 
and $\Tef = 4750$~K (HD72324), and 3783 K (HD146051). These four RGs have solar
metallicity, similar gravities, at least within the error bars, and differ by 
their \Tef. Visual inspection 
of Fig.~\ref{_templ} shows that the effective temperature of the secondary of \RS\
in 2006 September seems to be $4500\pm200$~K, if we ignore the 
presence of the molecular bands (such as TiO) in the RG spectrum.
Such molecular bands may be formed well way from the photosphere, in the cool
{\it molsphere} \citep{tsuj96} existing around RGs.
In the case of a complex system such as \RS\ we should regard the use of spectral features
for spectral classification with some caution:
the shape of the observed SED is more reliable for determining the effective
temperature of the RG. With this caveat,
the overall shape of the {\em continuum} of \RS, as well as some absorption features, 
agree best with the spectrum of HD72184. The best fitting SED template to the \RS\ spectrum, using $\chi^2$ criteria, confirm
this finding, see right panel of Fig. \ref{_templ}.
On the other hand, the molecular bands
in \RS\ are in better agreement with those in the spectrum of the M-giant HD146051. 
We revisit the case of \RS\ in Section \ref{cmod}. 

We acknowledge that our comparison of the continuum SEDs of \RS\
 and the RGs to estimate \Tef, as in Fig.~\ref{_templ}, is very
qualitative, because the gravities of the RGs
are not well known: a more accurate comparison would
compare stars having the same \logg, so that the RGs are at the
same evolutionary stage as the RG in \RS.
Furthermore we should note that the evolution of the secondary in \RS,
and the physical state of its atmosphere, differs
considerably from that of a ``normal isolated'' RG.
However, our simple and phenomenological comparison is model independent.
Therefore we use these results only for the purposes
of preliminary evaluation: more robust fits of our
theoretical spectra to the observed spectra will be
carried out in the Section~\ref{cmod}.

 The data in Fig.~\ref{_templ} show several emission lines superimposed on the
spectrum of the RG in \RS\  in 2006 September, and it is also clear that there is
a  hot component that contributes significantly below 4000\AA.
It seems likely that this excess is due
to the presence of the re-established AD and indeed
our attempt to model this part of the spectrum (see below) appears to confirm this;
however we shall see that fitting the SEDs below 4000\AA\ is possible for a wide range of
parameters for the newly-reformed AD. A full analysis is beyond the scope of
this paper; in general, it is not straight-forward to model this flux excess
using only the short wavelength range available at the edge of the observed
spectral region. In any case, this excess affects only a narrow spectral region,
and to simplify our discussion in the sense of reducing the number of the 
adopted parameters, we did not take into account the SED at these wavelengths
in our analysis. We believe that at
wavelengths $>4000$\AA\ {at least the atomic} absorption spectrum of \RS\ is formed mainly in
the photosphere of the RG.

\subsection{The spectrum at later dates}
\label{laterdates}
The observed SEDs of \RS\ at later dates differ considerably
from that in 2006 September. In the bottom panel of Fig.~\ref{_seds}
we compare observed fluxes in 2006 and  at later times. In general,
at the later dates, the emission features are stronger, and the continuum is
stronger in the blue; the latter is due to the development of the 
hot AD after 2006 September \citep{Worters07}.
Evidently the
contribution of the AD to the observed SED is significant at later 
dates. As a result, the overall slopes of the observed SED change: the
observed fluxes increase  to the red in 2006; on later dates the observed 
fluxes increase to the blue.

\begin{figure}
   \centering  
   \includegraphics[width=80mm]{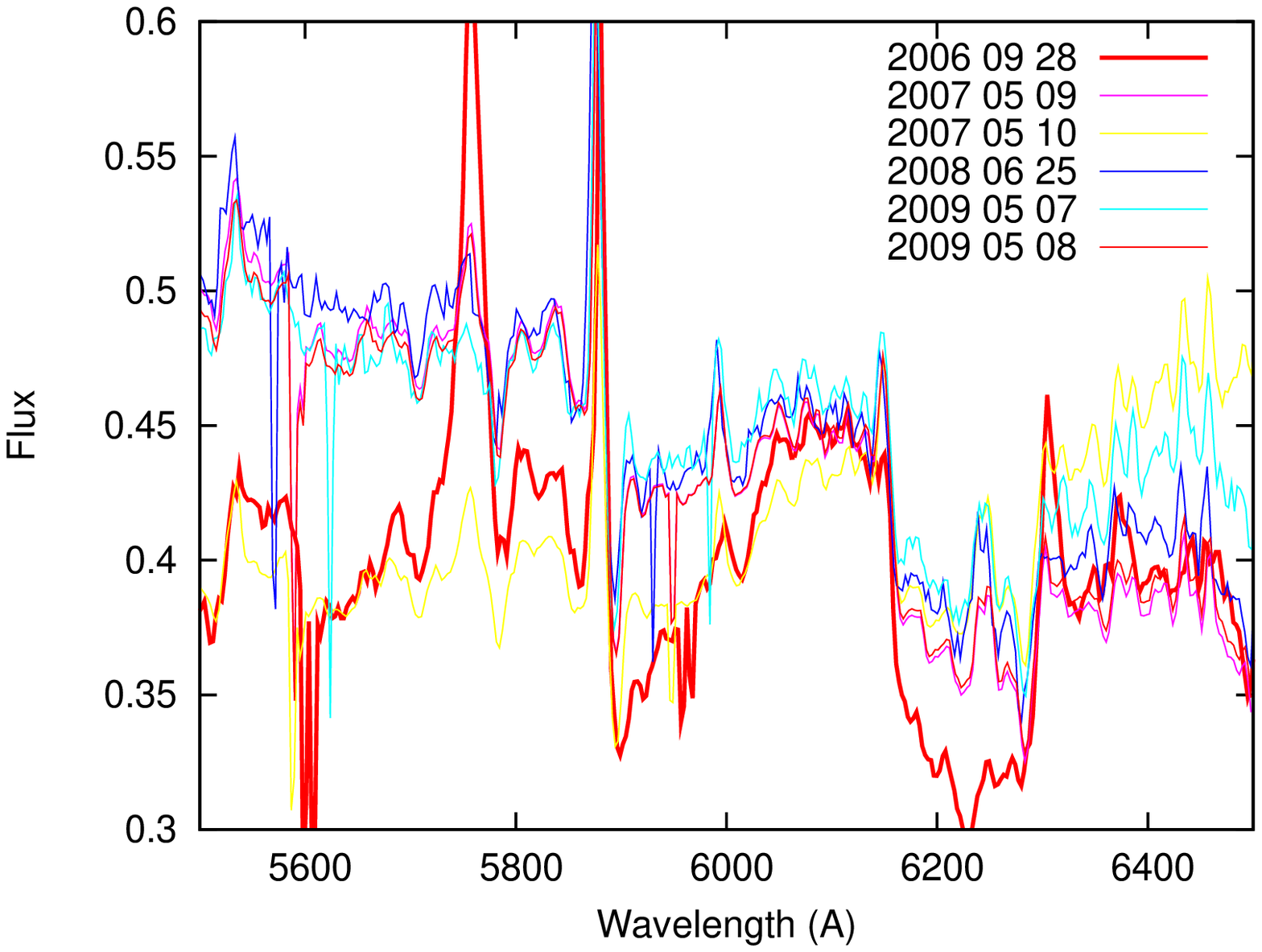}
     \includegraphics[width=80mm]{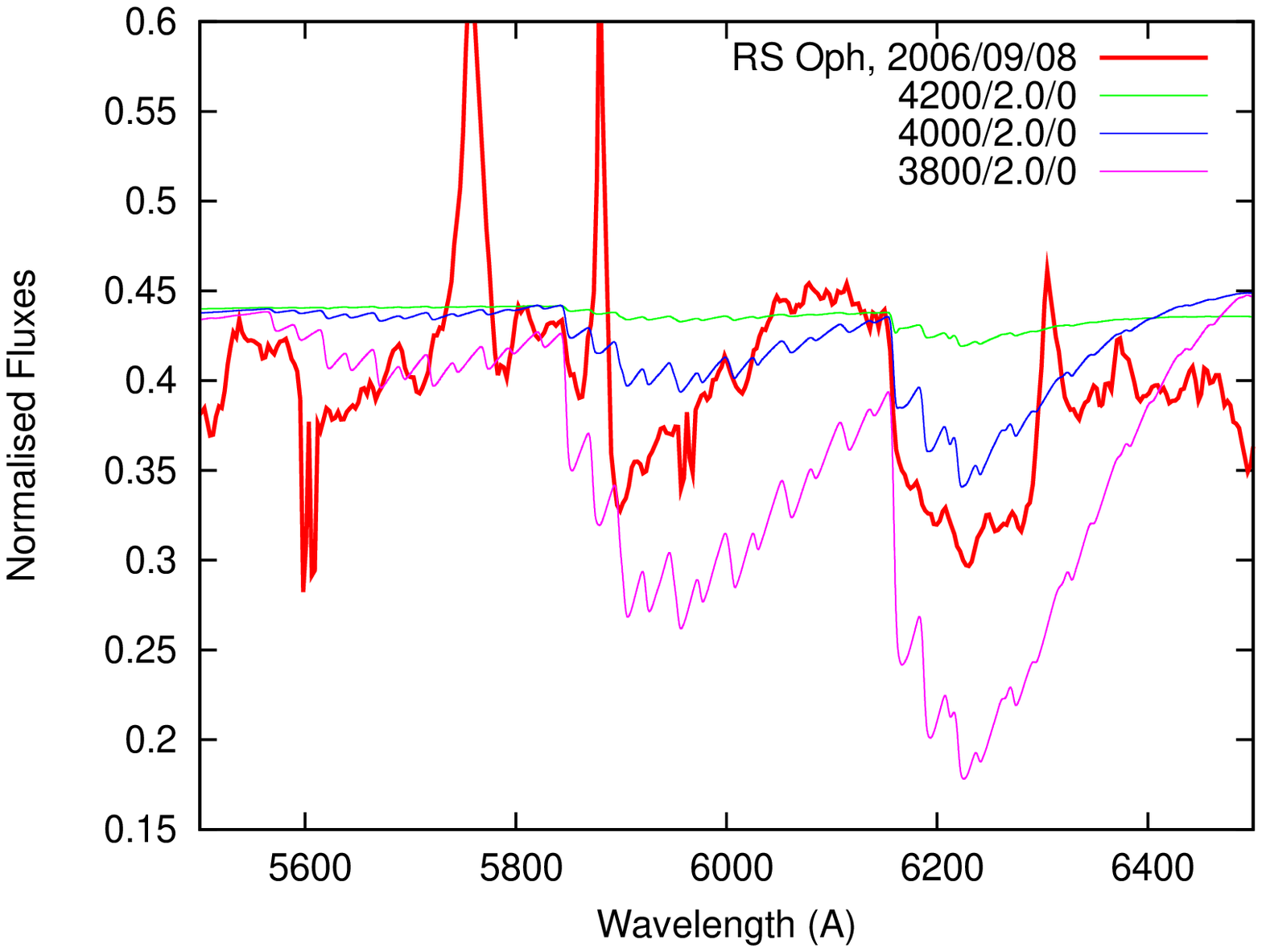}
      \caption{\label{_6200}  
      Top: the relative strength of the  6200\AA\ absorption in \RS\ 
      on the dates indicated. 
Bottom: comparison of detail in the spectrum of \RS\ around 6200\AA\
in 2006, with computed model atmospheres having different \Tef, 
showing profiles of the TiO $\gamma'$ band system across this spectral region. 
Theoretical spectra are convoluted with a Gaussian with $R = 1200$.
}
\end{figure}

A feature at $6160-6290$\AA\ is clearly seen in the spectra of \RS\ for all
dates. This is most prominent in the 2006 spectrum, after which its strength
decreases (see upper panel of Fig. \ref{_6200}). Likely this is
absorption in the TiO $\gamma'$ system (Fig.~\ref{_6200} lower panel), 
which has a band head at 6160\,\AA\ and which is known to be
the strongest band system in M-star spectra.
Interestingly, this feature is weak or even absent in the observed spectra 
of stars hotter
than 4300~K (see Fig.~\ref{_templ}). On the other hand,  our
theoretical modelling  
predicts the appearance of significant TiO spectral features  
only for model atmospheres having $\Tef < 4200$~K 
(Fig.~\ref{_6200} lower panel).
The appearance of TiO absorption features, despite being
reduced in strength by AD veiling after 2006, indicates
the presence of cooler regions in the RG atmosphere (see Section~4.3).

   \begin{figure}
   \centering
   \includegraphics[width=80mm]{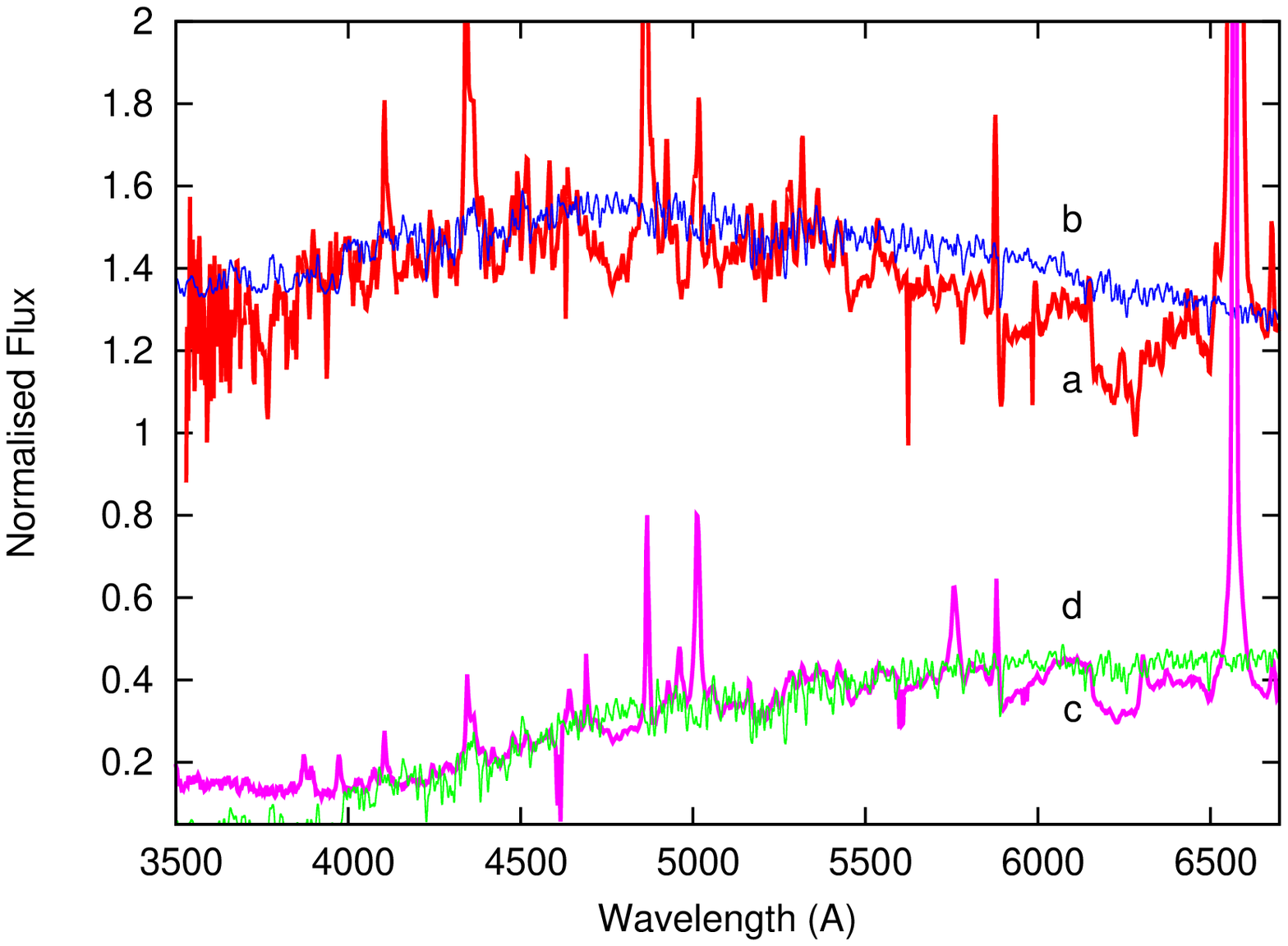}
   \includegraphics[width=80mm]{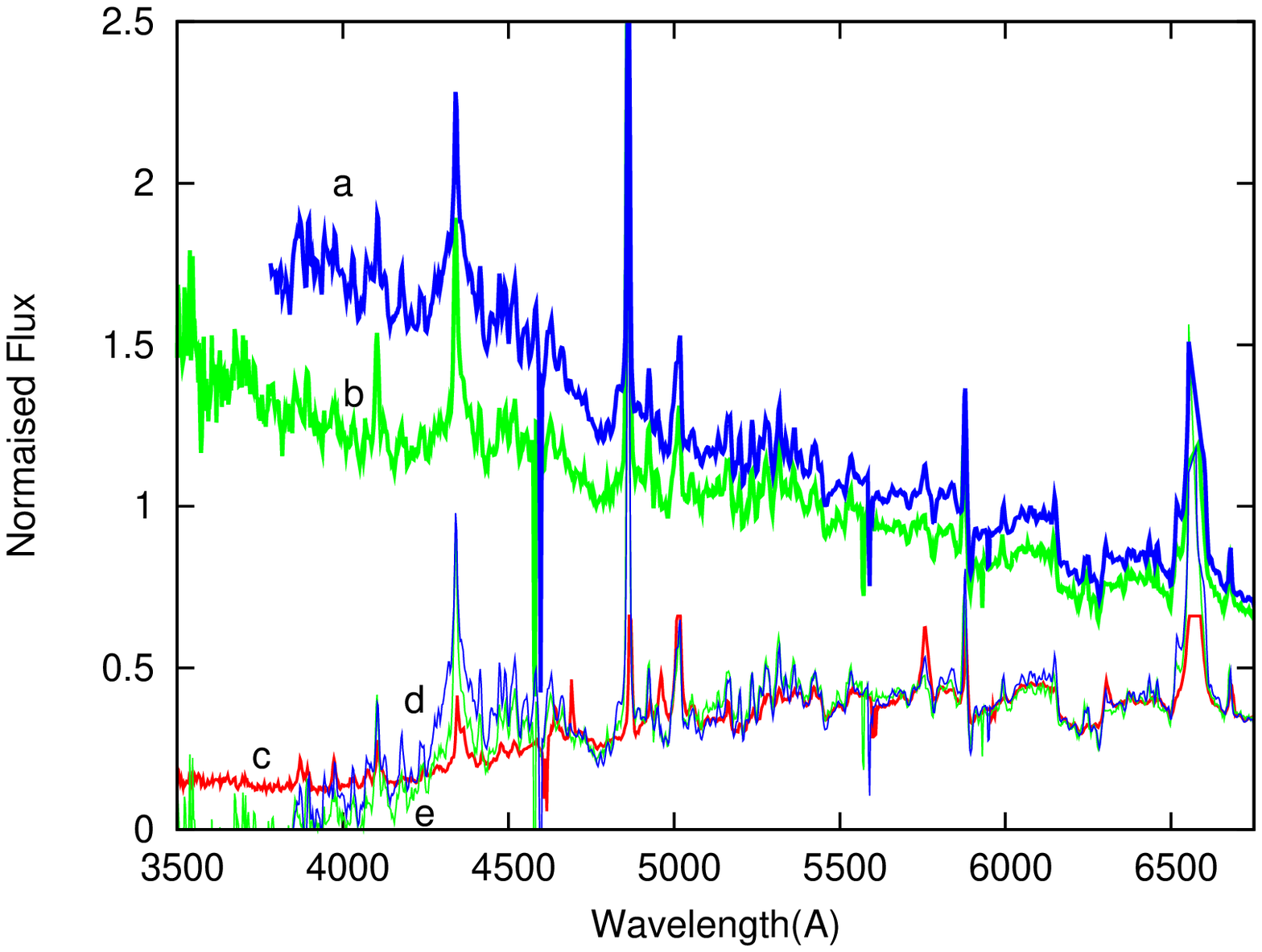}
     \includegraphics[width=80mm]{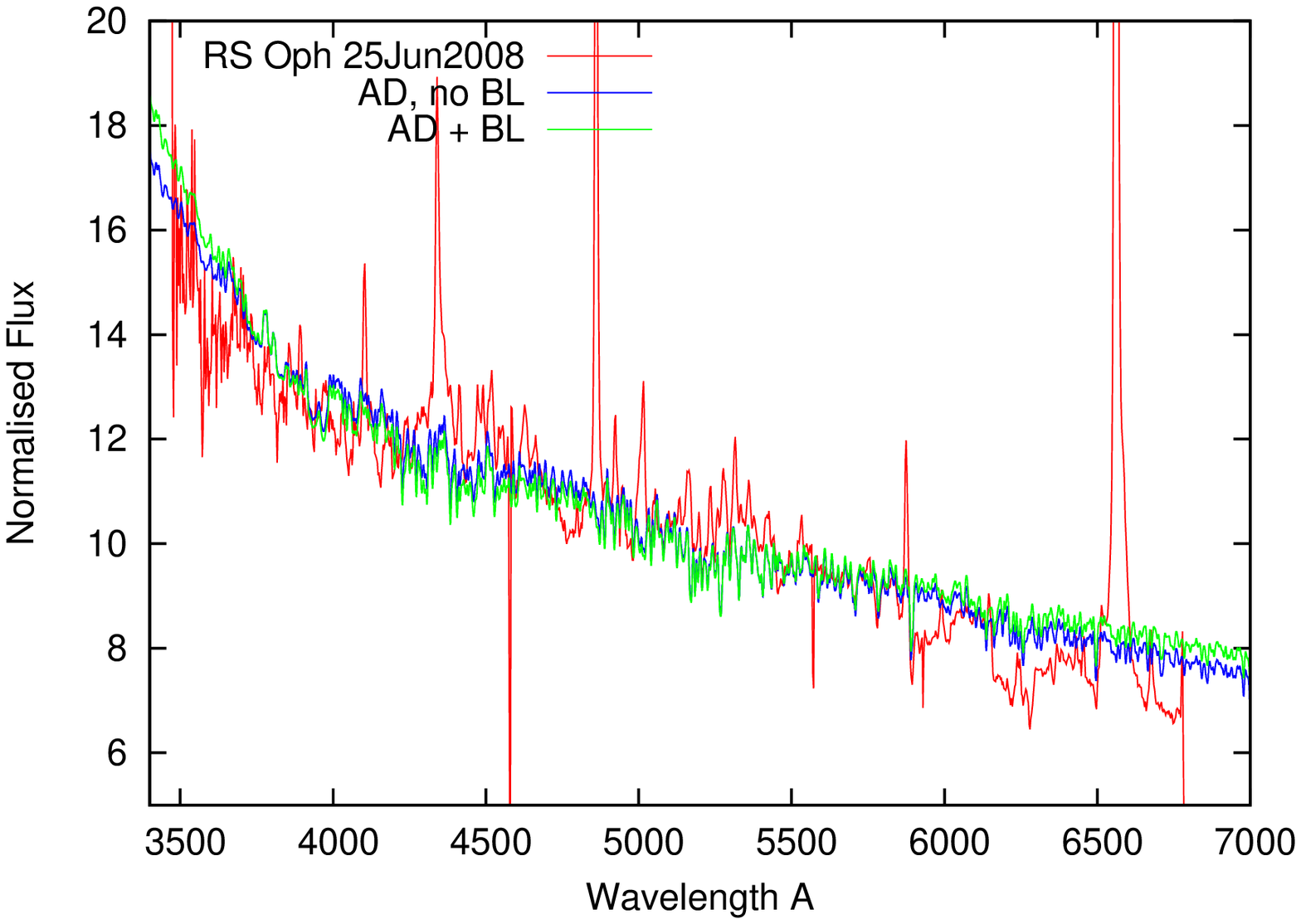}
      \caption{\label{_b1} 
Top: comparison of the SED observed on 2009  May 07 (a) with a model 
spectrum RG+AD (b), and extracted spectrum (d) with 9 Sept. 2006 (c).
Parameters of the RG model atmosphere and AD are  
$\Tef = 4200$~K, $\log g =2.0$, and $T_* = 20\,000 \pm 3\,000$, 
respectively.      
Middle: the observed spectra of RS Oph in 2006 Sept. 09 (c), 
2008 June 25 (b) and 2009  May 08 (a),
and the spectra of 2008 June 25 (e) and 2009 May 08 (d) 
after the extraction of the AD contribution.
Bottom: comparison of the SED observed in 2008 June 25 (thick line) with 
theoretical fluxes computed for our accretion model discs with and without BL.}
\end{figure} 

\subsection{Modelling spectra with an accretion disc \label{_ead}}

\subsubsection{A simple model including an accretion disc}

 First of all, to reproduce the observed spectra
after 2006, we add to
spectra obtained on 2006 September 28 the fluxes of
an AD, using the simple relation
\begin{equation}
F_{\rm RG + AD}=F_{\rm RG}+k*F_{\rm AD}(T_*),
\label{A-D}
\end{equation}
where $k$ is a scaling parameter. For simplicity we refer to this model of
the accretion disc as AD1.
For the flux distribution of the disc $F_{\rm AD}$ we use a standard  
optically thick disc which 
radiates locally as a black body \citep[e.g.][]{FKR}. The wavelength of
maximum emission depends on the
maximum disc temperature, which is given by $T_{\rm max}=0.49T_*$, and where the 
characteristic temperature $T_*$ is related to the accretion 
rate ($T_*\propto[\dot{M}\rho]^{1/4}$, $\rho$ being the density 
of the WD \citep[see][]{FKR}).

Various values of $\dot{M}$ for \RS\ have been reported in
the literature, ranging from e.g. 
$\sim2\times10^{-8}$\Mdot\ \citep[][based on X-ray
observations of \RS\ $\sim540-740$~days after the 2006 eruption]{_nelson11}
to $\sim3.9\times10^{-6}$\Mdot\ \citep[]
[a paper that summarises literature values of $\dot{M}$ for \RS]{Schaefer09}.
As the WD is close to the
Chandrasekhar limit, there is little scope for variation in $\rho$, so
this range of $\dot{M}$ translates to a factor $\ltsimeq4$ in the range of $T_*$.
The inner and outer disc radii
\Ras and \Rd were varied to obtain the fit to the observed SEDs.

To factor in the AD we first carried out fits to the data
for the later dates, using the lower accretion rates reported for \RS, namely
$\dot{M} \sim2\times10^{-8}$\Mdot.
Our numerical modelling shows that the low $\dot{M}$ values
correspond to the SED that shows a drop of flux in 
the blue part of spectrum of \RS\, such as we see only for  2009 May 7, 
see Fig. \ref{_b1}.
Fits to the observed fluxes in the blue
spectrum of \RS\ can be obtained with $T_*=20\,000$~K 
for 2009  May 7; the solution with $\dot{M} \sim2\times10^{-8}$\Mdot\
with the parameters listed in Table \ref{_TBL}. 

However, observed SEDs for other dates do not show the same drop in the
blue part of the spectrum: they seem to require a higher $T_*$, corresponding
to a higher $\dot{M}$. For our present purposes we adopt $T_*=68\,000$~K, 
yielding a maximum disc temperature $T_{\rm max} \simeq 33\,300$~K and
which corresponds to the higher accretion rate,  $\sim3.9\times10^{-6}$\Mdot.
Indeed for this accretion rate
we obtain fits of similar quality for the other four dates
(see Table~\ref{_TT}).
This procedure allows us to extract the contribution of the AD from the
five  spectra that appear to be affected by the presence of the disc.
We refer to these spectra as ``AD-extracted''.

\begin{table}
\caption{Parameters \Ras and \Rd of the AD1 model, in \Rsun, 
here $\dot{M} = \sim3.9\times10^{-6}$.
\label{_TT}}
\centering
\begin{tabular}{cccc}
\hline
Date         &  $T_*$ (K)   &   \Ras      &  \Rd  \\ \hline
09 May 2007  &   68000     &   0.5  $\pm$ 0.1        &   7.2 $\pm$ 1.5  \\
10 May 2007  &   68000     &   0.42 $\pm$0.08        &   6.5 $\pm$ 1.5  \\
25 Jun 2008  &   68000     &   0.42 $\pm$0.08        &   6.5 $\pm$ 1.5  \\
07 May 2009  &   20000     &   1.0  $\pm$ 0.1        &   7.5 $\pm$ 1.5  \\
08 May 2009  &   68000     &   0.76 $\pm$0.15        &   8.0 $\pm$ 1.5  \\
\hline\hline
\end{tabular}
\end{table}

\begin{table}
\caption{Parameters \Ras and \Rd of the AD2 model, in \Rsun, here 
$\dot{M} = \sim2\times10^{-8}$. We adopted for all dates
 $T_*$ = 20000 K, \Ras =  1.0  $\pm$0.2, \Rd =  7.5 $\pm$ 1.5
 \label{_TBL} }
\centering
\begin{tabular}{ccc}
\hline
Date       &  $T_{BL}$ (K)   &  $f\times 10^{-9}$   \\ \hline 
             &                 &       \\
09 May 2007  &   200\,000      &  10      \\
10 May 2007  &   200\,000      &   4      \\
25 Jun 2008  &   200\,000      &   4      \\
07 May 2009  &   no BL       &  no BL    \\
08 May 2009  &   200\,000      &  6.5     \\
            &                        &        \\
\hline\hline
\end{tabular}
\end{table}

There is general agreement for the parameters  \Ras\ and \Rd\ 
derived from the fits to the observed
spectra using the simple AD model for
2007 May 10 and 2008 Jun 25 (see Table~\ref{_TT});
in view of the fact that all the spectra were obtained close to
 quadrature (see Table~\ref{tab:bokobslog}) this is perhaps not unsurprising.
We note that, in this model,  \Ras\ and \Rd\
depend on the normalisation constant used to fit theoretical 
spectra to the observed SEDs. We also note that 
the values of \Ras\ and \Rd\ given in Table~\ref{_TT} 
are for normal orientation of the orbital plane.
Taking into account the orbital inclination of \RS, $i =39^\circ$
\citep{ribeiro}, we obtain more realistic 
values for the outer radius \Rd\ of the disc of $\simeq 10$\Rsun\ and $\simeq12$\Rsun.
These agree within a factor $\sim3$ of theoretical estimates by
\cite{wynn}, \cite{_king2009} and more recently \cite{alexander2011}, who predict that 
the Roche lobe in \RS\ may extend to 25 \Rsun, and that the radius of the AD
should be $\sim17$\Rsun.

\subsubsection{A two component model for the hot source}

In general, the one-component model of AD contribution allows us to obtain a formal 
solution, but with too high a $T_*$ and therefore too high an accretion rate.
 In the AD1 model, a higher accretion rate leads to a hotter AD.  
We have therefore explored a second model with the lower accretion rate
$\dot{M} \sim2\times10^{-8}$\Mdot \citep{_nelson11},
taking into account the flux from in RG and the AD, but now we factor in
a hot boundary layer (BL) at the inner boundary of the AD.
We refer to this model of the AD plus BL as AD2. 
To compute the total flux we use
\begin{equation}
F_{\rm RG + AD + BL}=F_{\rm RG}+k \times F_{\rm AD} + f \times F_{\rm BL} \:.
\label{A-DD}
\end{equation}
The factors $k$ and $f$ are determined from the fit to the observations;
they depend on the distance to the system and on the column density of
the emitting plasma. As in the one-component model AD1,
$F_{\rm AD}$ is a function of $T_*$, \Ras and \Rdp.
This two-component model for the hot source has the added bonus
that  the observed rapid variability may be explained mainly
by instability in the BL rather than in the entire disc, the
luminosity of which remains roughly constant.

Our modelling suggests that the spectrum for 2009  May 7 seems not to
have any BL contribution, or this contribution is very weak. For the other spectra
we adopt the same $T_{*}$ as for 2009 May 7, namely 20\,000~K,
which corresponds to lower accretion rate.  Following \cite{FKR}, \TBL =5 $\div$ 10 $T_*$;
for all our cases we found best fits for $T_{\rm BL} = 10T_*$ = 200\,000~K. 
In this case all the other spectra can be fitted by varying only the contribution of the BL,
but with reducing $k$  by the factor $\sim$ 3. For the AD2 model this
can be accounted for by changes in column density of the emitting  material.
Note that $f$ seems to change by a factor 2.5 on a time scale of one day
(see Table~\ref{_TBL}), if the changes in the observed SEDs are real.
Of course other solutions with higher $T_*$ and lower BL contribution
may also be possible. 

In the lower panel of Fig. \ref{_b1} we compare fits 
of our theoretical spectra, computed with AD1 and AD2, and different accretion rates
to the spectrum observed on  2008 June 25.
We find rather marginal differences for fits by models with 
different accretion rates. The models are similar in their representation of the observed fluxes and provide physically plausible solutions.
To provide a more conclusive result, we adopt a more sophisticated approach, modelling the SEDs in a broader spectral range.

\section{Modelling observed SEDs of the secondary}

\subsection{Fit to the optical spectrum of \RS \label{cmod}}

We found that the observed optical SEDs in 2007 -- 2009, 
  after the extraction of the AD contribution, 
  are very similar to the AD-free spectrum observed in
Sept. 2006, except for the emission lines which change
in intensity; see Fig.~\ref{_b1} for the dates in 2008, 2009.
This provides strong confirmation of our suggestion that
the physical state of the RG atmosphere does not change during 2006--2009.
Noting that the removal of AD is highly model-dependent, we use in
our further analysis the optical spectrum obtained in 2006,
which is minimally affected by AD emission:
analysis of the 2006 spectrum should provide the most satisfactory result 
because the data were obtained at a time when the AD was weak or absent.

We determined the effective temperature of the RG 
by fitting theoretical fluxes to the observed SEDs in 2006 September.
A small grid of classical model atmospheres in the \Tef\ range 3500--4700~K with
steps of 100~K, $\log g = 0, 1, 2$, and  abundances obtained in 
\cite{pavlenko08_rsoph} was computed with the SAM12 program \citep{pavlenko03}. 
Details of the atomic and  molecular line lists 
are described in \cite{pavlenko08_rsoph,pavlenko10}.
Atomic line parameters are from VALD \citep{kupka99}, 
and the molecular lines of TiO, MgH, CN are from the line lists of
\cite{plez98}, \cite{kurucz93}, and \cite{yakovina11}, respectively. Abundances
of carbon and nitrogen were taken from \cite{pavlenko08_rsoph}.
We adopted a microturbulent velocity of \Vt = 3.0\vunit. 
For the model atmospheres we computed synthetic spectra in the LTE approximation
using the WITA618 program \citep{pavlenko97}.
 
 Our computations were carried out for a wide range of effective temperatures 
and gravities. Strictly, \logg$ = 1 - 2$, the range of \logg\
for the RG, while \logg = 0 relates more to luminosity class II.
On the other hand, the physical state of the atmosphere of the RG in \RS\
differs from that of an isolated ``normal'' M-giant as the RG in this case is the
secondary in an interacting binary system, with the resultant effects of
irradiation, tidal distortion, strong stellar wind etc. The parameter \logg\
should be regarded as an ``effective'' gravity, averaged over the time of observations;
``gravity'' in the conventional sense cannot be used to describe gravitational 
properties in dynamically changing system. Fortunately, the orbital period of \RS\ 
is sufficiently long that quasi-stationary equilibrium is established in the RG atmosphere,
at least at the level of its photosphere. 

Our synthetic spectra, smoothed by a Gaussian corresponding to $R=1\,200$,
were fit to the observed SED of \RS\ in 2006.
Fits were done for all synthetic spectra
in our grid, but omitting emission lines as they form in the 
circumstellar environment
and outside the photosphere of the
secondary. The best 
fit was determined by minimising the parameter
$$S(f_{\rm h}) =
   \sum_\nu \left ( F_{\nu} - f_h \times F_{\nu}^x \right )^2  , $$
where $F_{\nu}$ and $F_{\nu}^x$ are the fluxes in the observed and computed spectra,
respectively, and $f_{\rm h}$ is the normalisation factor. A similar procedure 
was used by \cite{pavlenko08_rsoph,pavlenko10} to fit 
synthetic spectra to the observed spectra of \RS\ in the IR region.

   \begin{figure}
   \centering
   \includegraphics[width=80mm]{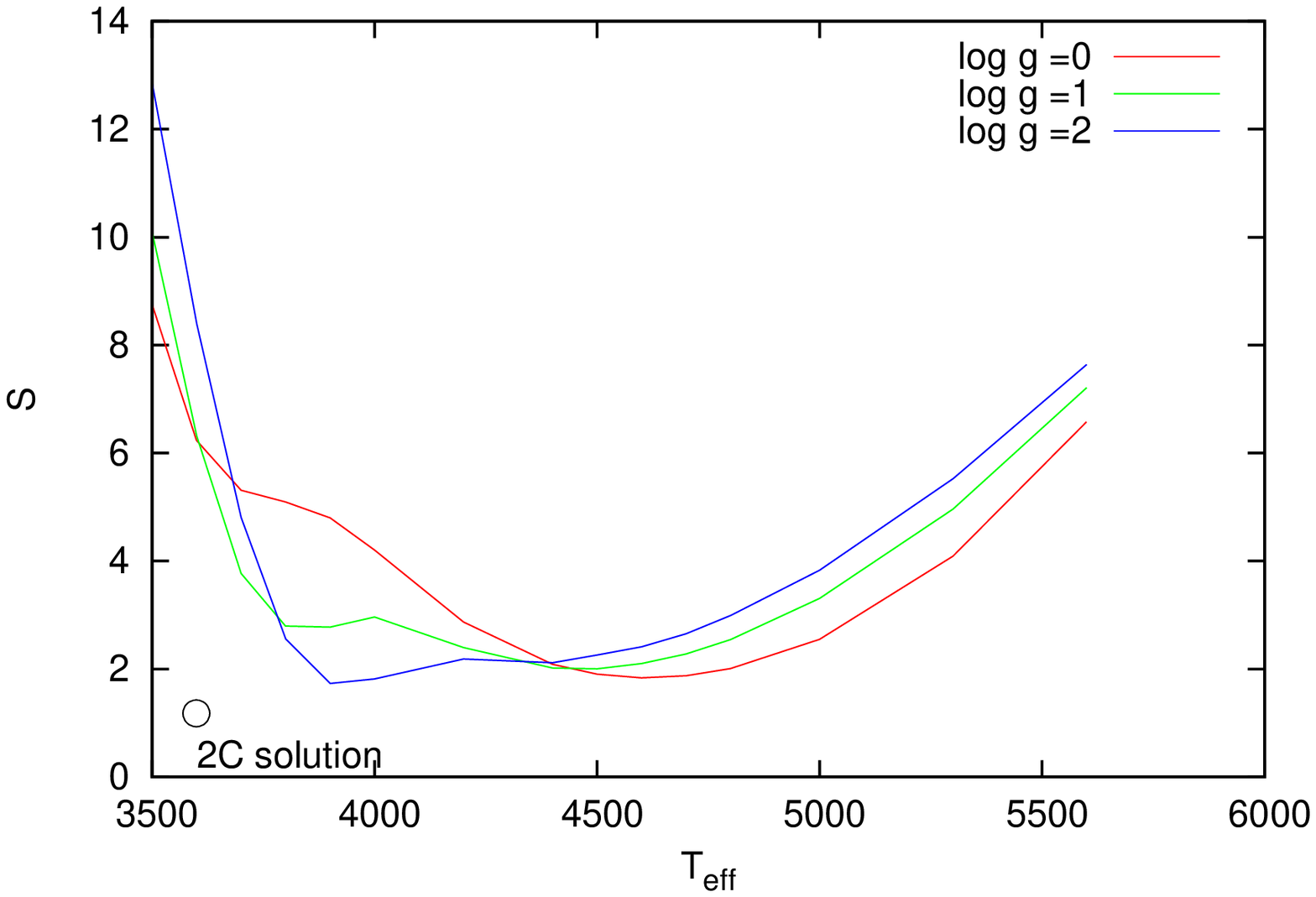}
     \includegraphics[width=80mm]{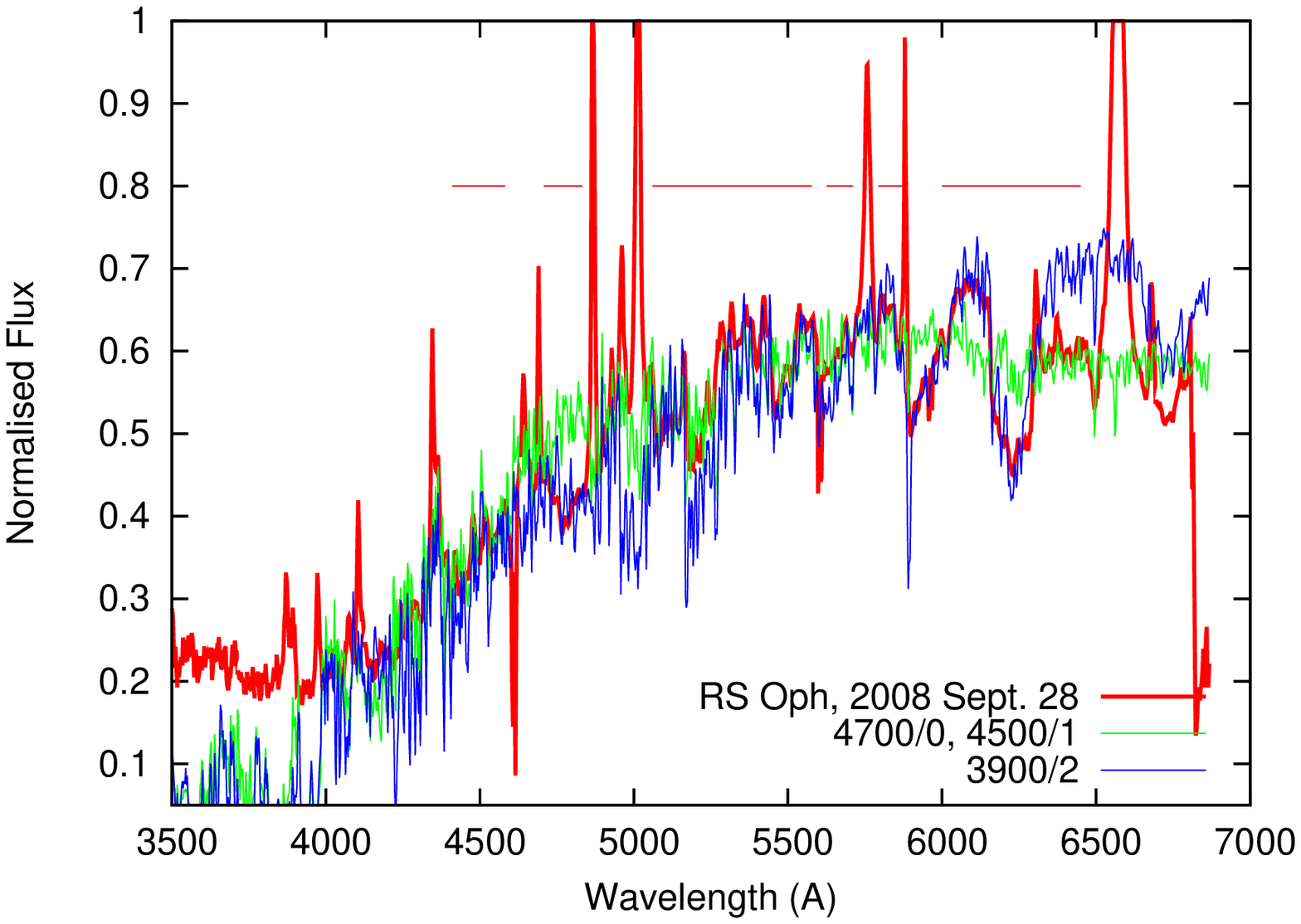}
      \caption{\label{_4500}  
     Top: minimisation parameter $S$ derived in the fitting
     of our synthetic spectra with model atmospheres of different 
     \Tef\ and $\log g$ to the SED of RS Oph observed in 2006.
 Circle on top panel shows solution (3600/0.0) found with 2C model 
     (see Section~\ref{_2c}).
Bottom: the best fit of synthetic spectra, computed with 
     a 4700/0.0, 4500/1.0 and 3900/2.0 model atmospheres to the 
 SED of RS Oph observed in 2006.
     Horizontal lines show the spectral regions used to compute $S$.
      Vertical scale arbitrary.}
   \end{figure}

The results of fitting our theoretical spectra to the spectrum 
of \RS\ in 2006 are unexpected.
Formally the best solution is for model atmospheres with \logg\ = 0.0 and 1.0, which provide 
well marked minimum $S = 1.836 \pm 0.002$ and $S = 2.004 \pm 0.002$    
at \Tef\ = 4600~K and 4500~K, respectively. In general, these results agree with
those we obtained from comparison of the observed SEDs of RS Oph with
template stars (see Section~\ref{template}). On the other hand, the best solution for model
atmospheres having \logg\ = 2.0 leads to a much lower \Tef, 3900~K
with $S = 1.731 \pm 0.002$. Furthermore,
a local minimum for $S$ at 3800~K is seen in the $S$--\Tef\ plot computed for
model atmospheres with \logg\ = 1.

Nevertheless, the formal fits to the observed SEDs (Fig.~\ref{_4500})
and comparison with the optical spectra of template stars (see Fig.~\ref{_templ})
provide a wide range of the effective temperature for the photosphere of
the RG in \RS,  $\Tef = 3900 - 4700$~K,
with uncertainties $\pm100$~K.
Our fits show that, by lowering $\log{g}$, we obtain 
higher \Tef\ due to the decreased strength of TiO bands in the models 
having lower \logg.

By using optical spectra we have obtained a noticeably 
different effective temperature for the \RS\ secondary to that
obtained from modelling the
IR spectra \citep[$4100\pm100$~K;][]{pavlenko08_rsoph}.

\subsection{Fit to the infrared spectrum of RS Oph
\label{__fitsir}}

In this section we analyse the IR spectrum of the RG, as
observed 2 years after the 2006 eruption, on 2008 July 15 
(see Section \ref{_irs}).
Within two years of the eruption, the photosphere and the structure 
of the convective envelope of the secondary had recovered,
and the AD had reformed into a quasi-stable state.

Consequently, for the 2008 data, we can expect a noticeable 
contribution from the AD to the entire \RS\ spectrum, even in the IR.
To remove the contribution of the AD
from the IR spectrum of \RS\ observed in 2008 July 15, we use the procedure
already described in Section~\ref{_ead};
these data will again be referred to as ``AD-extracted''.

As before, synthetic spectra for $1.2 < \lambda (\mic) < 2.5$ were computed, 
including absorption from TiO, H$_2$O, CO, CN
for the grid of atmospheres. These spectra were then fitted to the AD-extracted IR
spectrum of \RS\ on 2008 July 15.

\begin{figure}
   \centering
   \includegraphics[width=80mm]{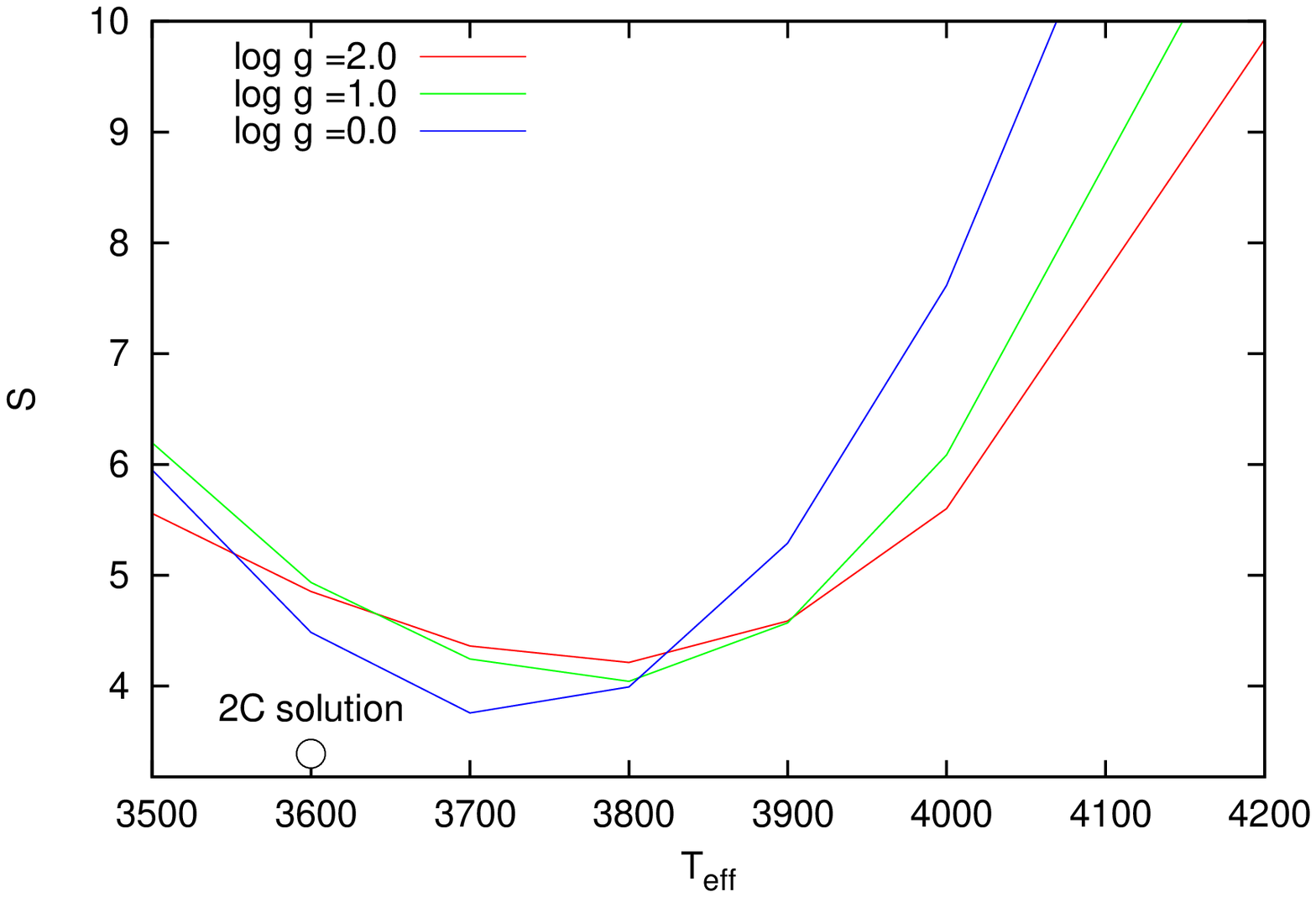}
     \includegraphics[width=80mm]{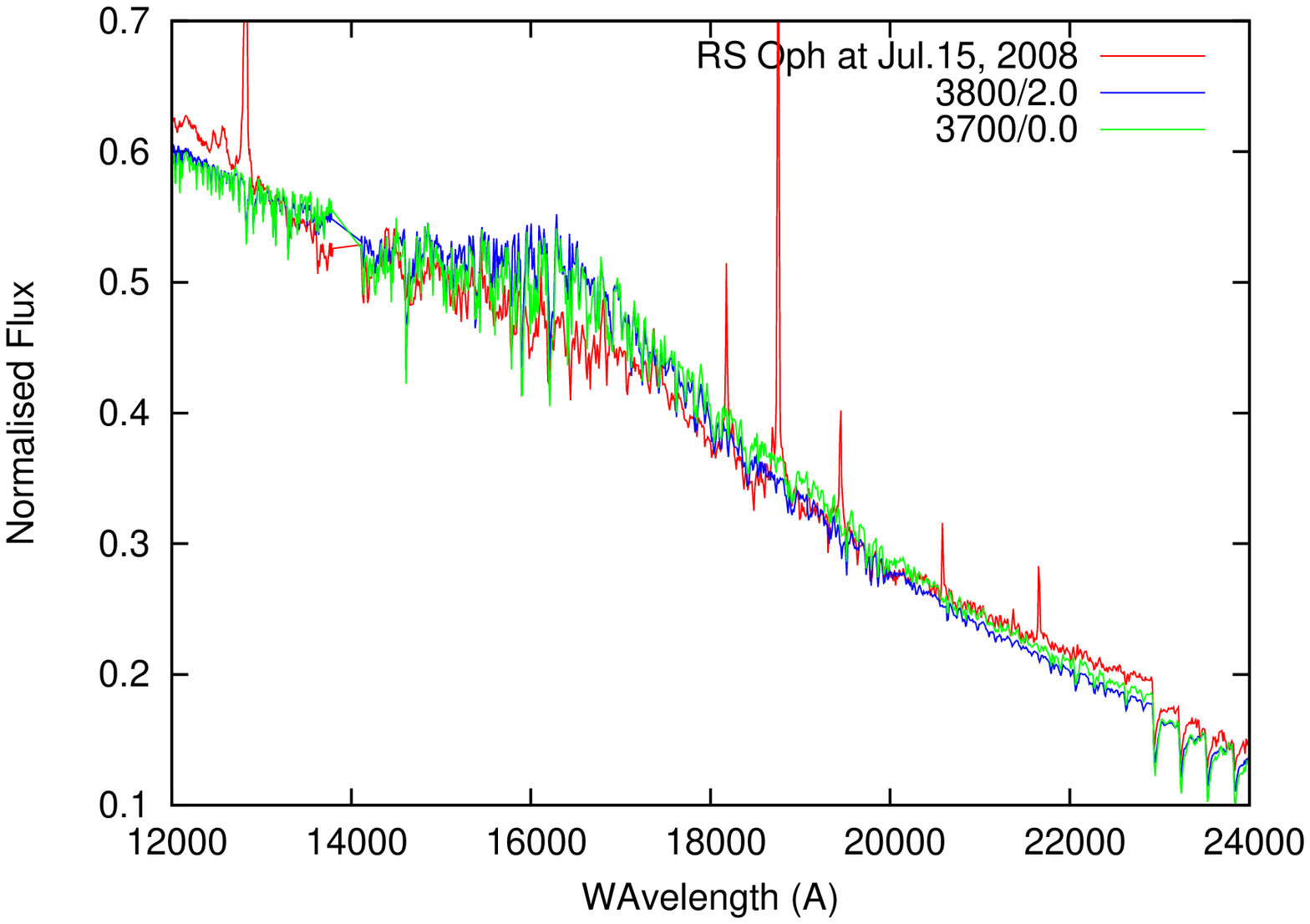}
          \includegraphics[width=80mm]{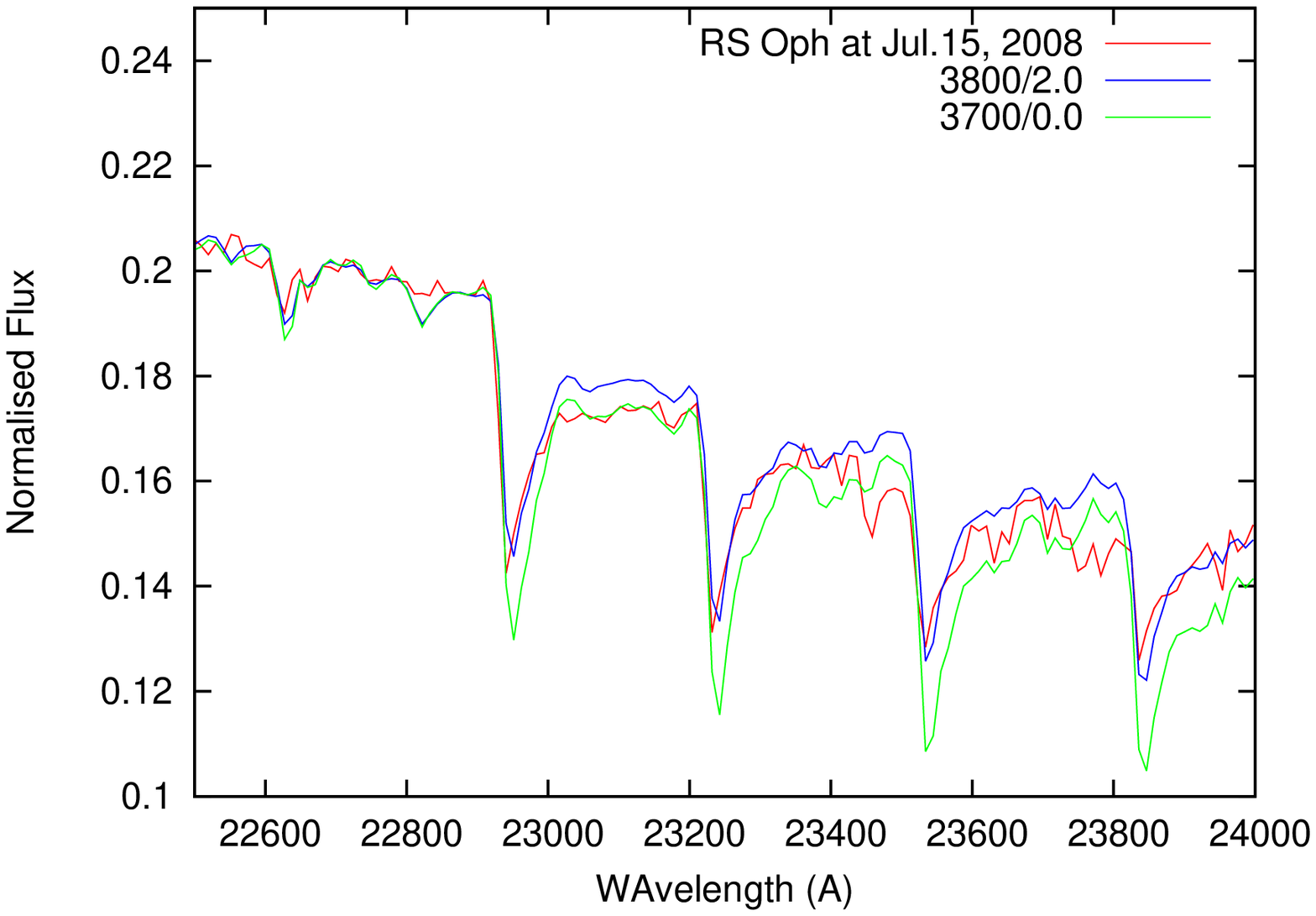}
      \caption{\label{_ir2} Top: dependence of $S$ on \Tef\ obtained by
      fitting synthetic spectra to the AD-extracted 
           IR spectrum of \RS\, from 2008 July, 15. 
     Circle on top panel shows solution (3600/0.0) found with two component model.
     (see section \ref{_2c}). 
     Middle: the best fitting
synthetic spectrum computed for a \Tef/$\log{g}$ 3800/2.0
and 3700/0.0 model atmosphere to the
non-extracted spectrum from 2008 July, 15. 
Bottom: fit of synthetic spectra computed for 
3800/2.0 and 3700/0.0 model atmospheres 
to CO molecular bands in the same observed spectra.
}
   \end{figure}

The dependence of the parameter $S$ on \Tef, obtained by fitting 
our synthetic spectra to the AD-extracted IR spectrum of \RS\ on 2008 July 15 
is shown in the upper panel of Fig. \ref{_ir2}.
In the middle panel, we show the best fit to the observed spectrum of RS Oph.
Our new fit to the observed and extracted IR spectrum was carried out over 
a broader wavelength region, at $1.2 < \lambda (\mic) < 2.3$, than in 
\cite{pavlenko08_rsoph}.
As before, the strong spectral features (such as emission lines or 
strong telluric bands) were excluded from consideration. 
The best fit to the AD-extracted IR spectrum results in lower temperatures,
3700--3800~K, for our adopted  $\log{g}$, than was obtained 
by \cite{pavlenko08_rsoph}; see Fig.~\ref{_ir2}.

The effective temperature of the secondary in \RS, obtained by fitting
the infrared spectral region, is consistent with the lower end of the
temperature range determined from the optical spectra. 
Moreover, a comparison of the observed and computed IR spectra for
\Tef/$\log{g}$ = 3800/2.0 in the spectral 
range containing the CO first overtone bands ($\lambda\lambda$ 2.28 -- 2.4 \mic)
shows good agreement (see the bottom panel of Fig. \ref{_ir2}).
This spectral range is severely affected by telluric absorption, and usually
only the first two
bands with heads at 2.29 and 2.32\mic\ can be used for analysis \citep[see][]{pavlenko03a,pavlenko10}.
Nevertheless, despite the comparatively 
low resolution of our observed spectrum at these wavelengths,
these results correspond well enough with \Tef/\logg = 3900/2.0 obtained from the fit
to the SED in the optical spectral region (Fig. \ref{_6200}).  
 We note that \Tef = 3900 K does not agree with a higher effective
temperature for the RG obtained from comparison of the observed optical 
spectra with template stars (see Fig. \ref{_templ}). 

On the other hand, our low-temperature model cannot fit the broad feature
around 1.6\mic, commonly attributed to the $H^-$ ion opacity at these wavelength; this
is more prevalent in the spectra of 
cooler stars: in K--G giants it is much weaker or absent.
In a sense the absence or weakness of this feature in the 	
observed IR spectrum of \RS\ supports the choice of a
higher \Tef\ for the red giant  (see fits in \cite{pavlenko08_rsoph}, where a higher \Tef was adopted). 
This is discussed further in the next section.

\subsection{A two-component model atmosphere of RG \label{_2c}}
 We have derived the effective temperature of the secondary star in \RS\ from
IR data, and noted discrepancies in our results obtained from the optical spectra. 
Although the shape of the continuum indicates a relatively high effective 
temperature, the TiO bands in the same dataset suggest a lower effective temperature. 
It is clearly desirable to reproduce the observed optical continuum and absorption spectrum 
by a single model. 

Firstly, we consider a hotter pseudo\-photosphere to account for the higher temperature
implied by the optical continuum. This could be due to heating of the 
outermost layers of the RG atmosphere by irradiation from the primary.
However, such a one-component model cannot account for the presence of the TiO
bands in the observed optical spectrum.
A possible explanation is that these bands might form in a cool envelope
outside the \RS\ system, such as the torus harboring the dust reported by \cite{evans07}.
However the inclination of the \RS\ system is unlikely to lead to the TiO absorption observed.
Also, the presence in the observed spectrum of
strong emission lines provide some evidence that the outermost layers
of the envelope consists of hot ionised plasma. We conclude that model
atmospheres with cool outer layers are not realistic and we explore a two-component
model such as might arise from irradiation of the RG by the WD.

Irradiation affects only the hemisphere facing the primary. Therefore, we would
expect to see at least two temperature components in the RG atmosphere:
a low temperature region, in which the IR emission and TiO absorption arise, and an 
irradiated, hotter region, where at least a part of the continuum flux in the optical 
is formed. The hot region of
the RG atmosphere is most likely located in a spot facing
the primary. In this region, the
TiO bands will be veiled by the hotter continuum, as seen in the spectrum of \RS.
Multi-component model atmospheres 
have been invoked to account for stellar spectra where
there are both hot and cool regions on the photosphere (see \citealt{Strassmeier1997}).

To explore this possibility, we model the contribution of the hotter component
to the total flux emitted by the RG atmosphere, with fluxes forming in hotter parts
i.e. at higher effective temperatures \Thp.  
The contributions of the hot component over the whole spectral 
range of interest are computed for a grid of 
synthetic spectra formed in the hotter part(s) of \RS\ photosphere of  effective temperature 
\Thp\ = 4400, 5000, 5300, 5600,  6000 and 8000~K.
Then we define the contributions of the undisturbed (cool) RG atmosphere and irradiated region
to the total flux as $a$ and $1 - a$, respectively, i.e. the total observed flux from the 
two-component atmosphere is 
$$F_{total} = a \:F_{RG}+(1-a)\:F_H \:\: .$$
Here $F_{\rm RG}$ and $F_{\rm H}$ define the flux from 
the ``normal'' and ``heated''
parts of  the RG atmosphere, respectively.
Computations were carried out for all
the model atmospheres in our grid, and additionally $a$ = 0.98, 
0.95, 0.9, 0.8, 0.7, 0.6.
This range of $a$ was selected to get the best solution within
the adopted grid.

The procedure is as follows:
\begin{enumerate}
 \itemsep=1mm
\item fit synthetic spectra to the IR AD-extracted spectra 
of \RS\ and determine the best solution;
\item [{then}]
\item fit the AD-extracted optical spectrum observed for 2006 September,
using the best fit parameters in the IR as the starting point. 
\end{enumerate}
Ideally we would have obtained spectra in the IR and optical 
spectral regions at the same time. However,  
the dates of observations of the optical and IR spectra differ by 
about 20 days. We note that this period is not significant compared to the orbital
period of the system in 453.6$^d$ \citep[e.g.][]{Brandi2009}, but 
short period variations are seen on occasion, as noted 
in Section~\ref{_ostv}. Nevertheless, the optical spectrum of 2008 July 15
is similar to that of 2006 September 9, after the AD has been extracted  (see the middle panel Fig. \ref{_b1}).

\begin{figure}
   \centering
   \includegraphics[width=80mm]{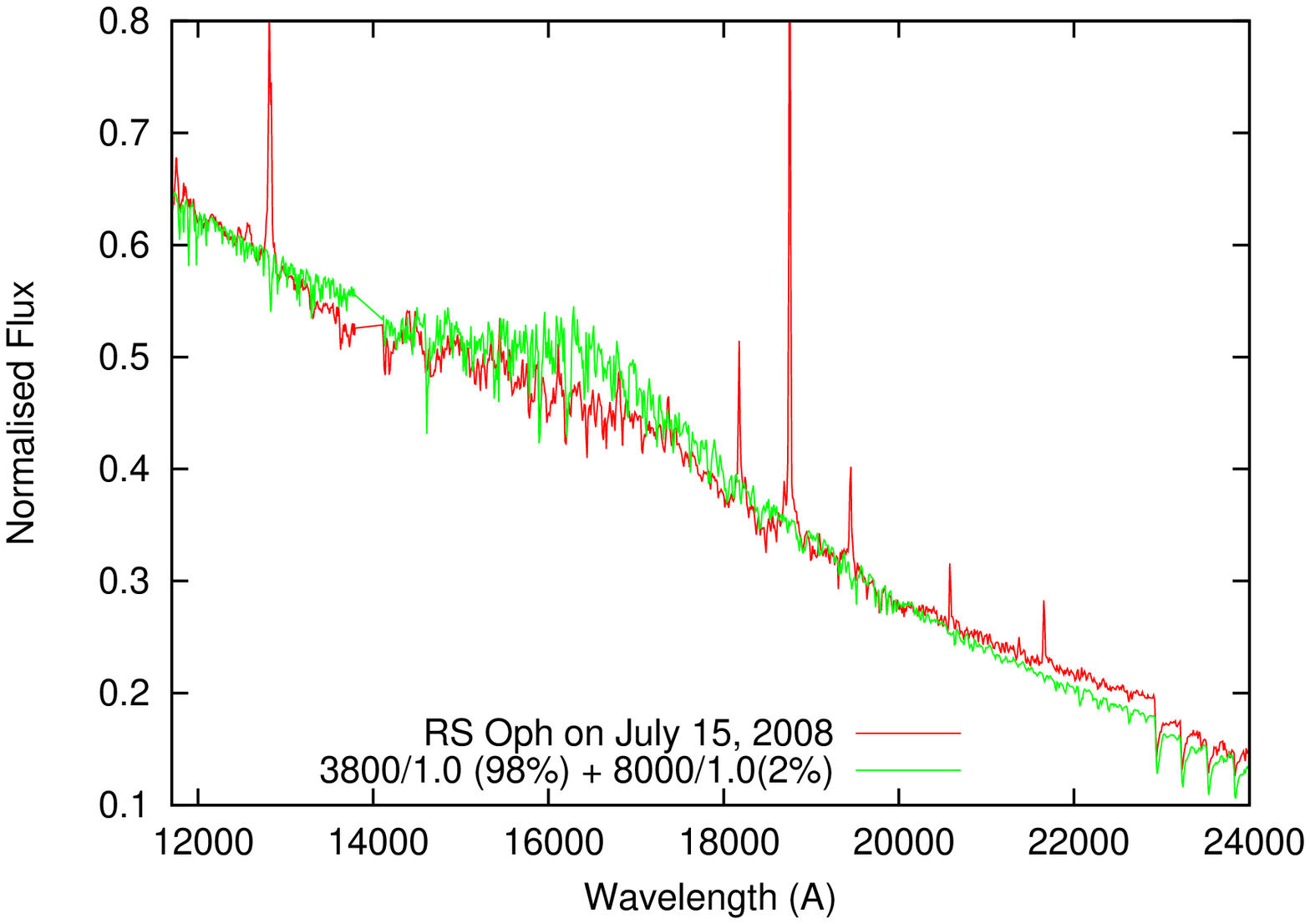}
     \includegraphics[width=80mm]{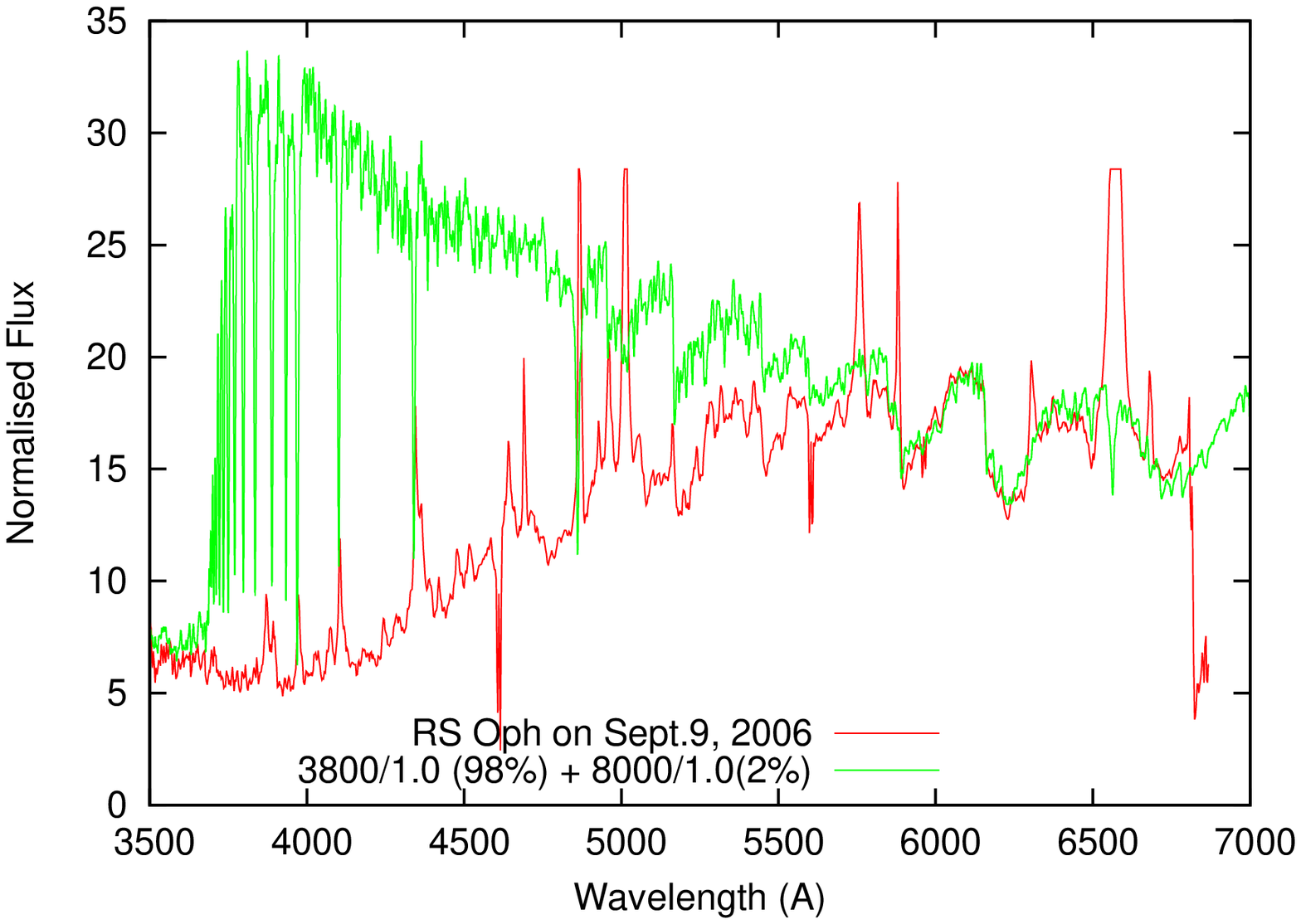}
      \caption{\label{_8000} Top: best fits to the AD-extracted IR spectrum 
of \RS\ on July 15, 2008
with the two-component model: (98\% with $\Tef/\log{g} = 3700/1.0$ and 2\% with 8000/1.0).
Bottom: fits of the AD-extracted optical spectrum on Sept. 9, 2006 with the same model.
 }
   \end{figure}

Initial numerical experiments clearly excluded the presence of regions
with  \Thp $ > 5800$~K in the RG atmosphere.
An example of our fits to the AD-extracted spectra of \RS, with a 
two component model having $\Tef = 3600$~K and $\Thp = 8000$~ K, is shown in the 
lower panel of Fig.~\ref{_8000}. 
Fits to the IR part of the spectrum seem satisfactory, with $a=98$\%, but 
the fit to the optical part of spectrum is very poor
(see bottom panel of Fig.\ref{_8000}).

Numerical modelling of our spectra using the two-temperature model 
allows us to determine a range of \Tef\ in the hotter part of photosphere,
$4400 < \Thp \mbox{(K)} < 5600$.
Within this \Thp\ range for the hotter atmospheric region, we obtain reasonable
fits to both optical and IR spectra, with the inclusion of the cooler component. 
For all two-temperature synthetic spectra, we compute the parameter $S$,
with minimum $S$ as usual giving the best fit to the 
AD-extracted fluxes in the IR spectra. 

The best fit to the secondary optical spectrum of 2006 September 26 was obtained
with $\Tef/\log{g}= 3600/0.0$ for 
90\% of the RG surface,  and \Thp/\logg = 5000/0.0  with $S$ = 3.131 $\pm$ 0.002 for 10\%
(see Table \ref{_resm}). In general the optical spectrum is much more sensitive to
the adopted \Tef; therefore we should regard this result with more confidence
in comparison with that obtained from the fits to IR spectrum. Nevertheless,
the same model provides one of the best fits to the \RS\ infrared spectrum with 
$S$ = 3.39 $\pm$ 0.002. We see better solutions for the IR spectrum
with lower \Tef\ for all \Thp,  in comparison with that found
from the fits to optical spectrum. Interestingly, most of our best solutions
of fits to optical and IR spectra 
were found for $a = 0.8 \div 0.98$ (see Table \ref{_resm}).   

We recall that both best fits to the optical and IR spectra of \RS\
for the two-component model atmosphere
were obtained with lower $S$ than for the classical model atmospheres
(see Figs. \ref{_4500} and \ref{_ir2}).

\begin{table}
\caption{Parameters of two component model atmospheres determined for the best fits
to the observed optical and IR spectrum in the format (\Tef/\logg : $a$) 
followed by the computed $S$ for the corresponding fits. Here \Thp
is the effective temperature of the heated part of \RS\ photosphere. Only 
three first the ``best'' solutions are given in the ascending order of $S$.\label{_resm}}
\begin{tabular}{cccc}
&&& \\
\multicolumn{4}{c}{Fits to the optical spectrum}                               \\\hline
                &                  &                  &                    \\
\Thp:~   4400 K       &      5000K       &    5300K         &   5600K         \\
                &                  &                  &                     \\
 3800/2.0: 0.7  &  3600/0.0: 0.9   & 3800/1.0: 0.95   & 3800/1.0: 0.95   \\
1.273 $\pm$ 0.002& 1.186 $\pm$ 0.002 & 1.221 $\pm$ 0.002 &  1.328 $\pm$ 0.002\\
                &                  &                  &                   \\
 3800/2.0: 0.8  &  3800/1.0: 0.9   & 3700/0.0:0.95    & 3700/0.0: 0.95 \\
1.297 $\pm$ 0.002& 1.193 $\pm$ 0.002 & 1.321 $\pm$ 0.002 &  1.344 $\pm$ 0.002 \\
                &                  &                  &                   \\
 3700/2.0:0.7   &  3700/1.0:1.20   & 3700/1.0: .95    & 3800/1.0: 0.98 \\
1.315 $\pm$ 0.002& 1.202 $\pm$ 0.002 & 1.378 $\pm$ 0.002 &  1.419 $\pm$ 0.002 \\
                 &&& \\
\hline
&&& \\

\multicolumn{4}{c}{Fits to the IR spectrum}                               \\\hline
                      &&& \\
\Thp:~   4400 K       &      5000K       &    5300K         &   5600K         \\
                &                  &                  &                     \\
3500/0.0: 0.8   & 3500/0.0: 0.9    & 3500/0.0: 0.9    &  3500/0.0: 0.9 \\
3.413 $\pm$0.002 & 3.131 $\pm$ 0.002 & 2.945 $\pm$ 0.002 &  3.086 $\pm$ 0.002 \\
                &                  &                  &                   \\
3500/1.0: 0.8   & 3500/1.0: 0.9    & 3600/0.0: 0.95   &  3600/0.0: 0.95  \\
3.588 $\pm$ 0.002& 3.364 $\pm$ 0.002 & 3.217 $\pm$ 0.002 & 3.150 $\pm$ 0.002 \\
                &                  &                  &                  \\
3600/0.0: 0.9   & 3600/0.0: 0.9    & 3500/1.0: 0.9    & 3500/0.0: 0.95 \\
3.586 $\pm$ 0.002& 3.390 $\pm$ 0.002  & 3.231 $\pm$ 0.002 & 3.167 $\pm$ 0.002 \\
            &                  &                  &                  \\\hline

\end{tabular}
\end{table}

On the other hand, we obtained these results by modelling 
optical and IR spectra from 2008 separated by 20 days; we believe that 
\RS\ was in a quiescent state over this period. Indeed the optical spectrum
on 2008 June 25 seems, after extraction of the AD, very similar to its state
in 2006, when there was no disc contribution.

\begin{figure}
   \centering
   \includegraphics[width=80mm]{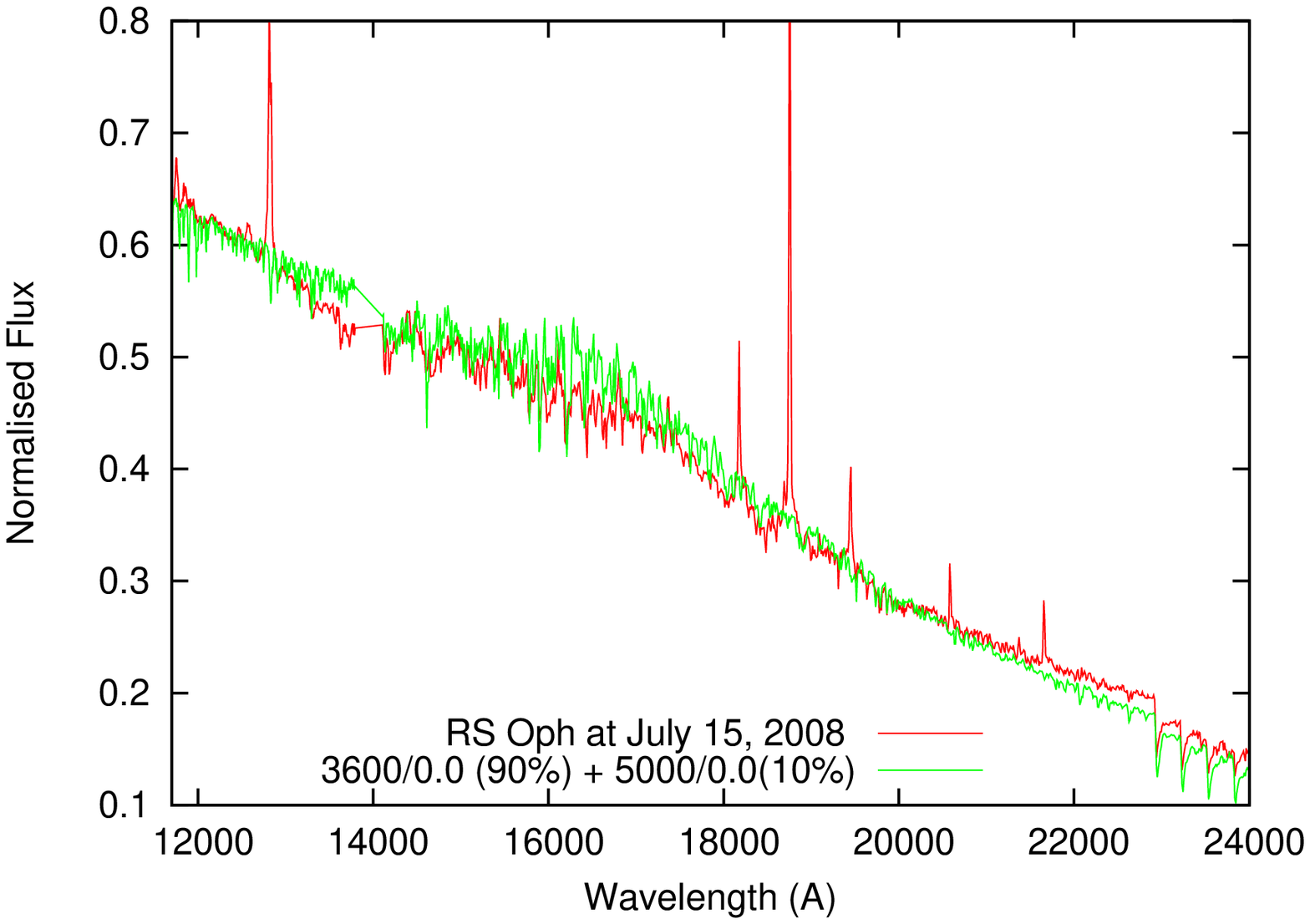}
     \includegraphics[width=80mm]{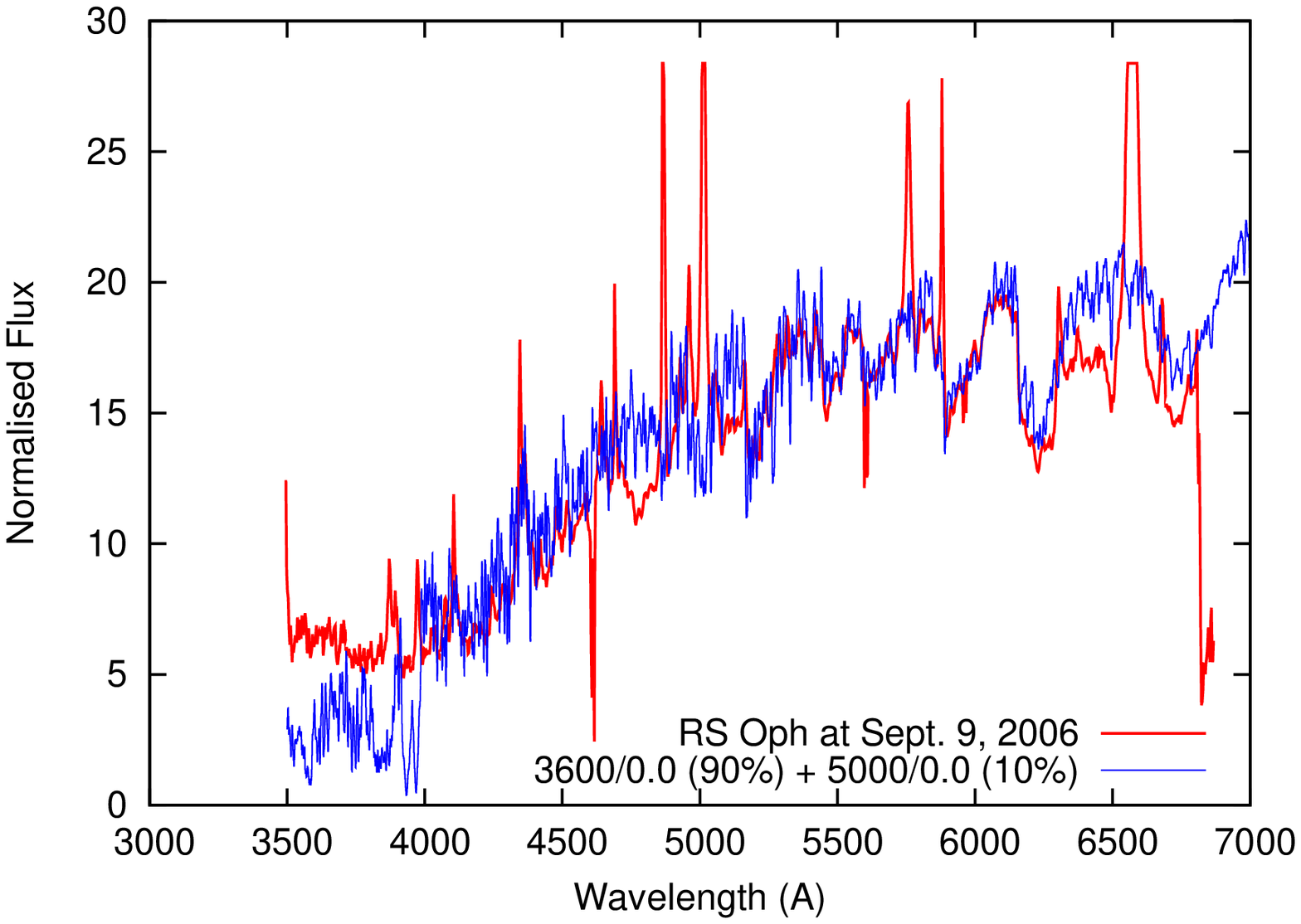}
     \includegraphics[width=80mm]{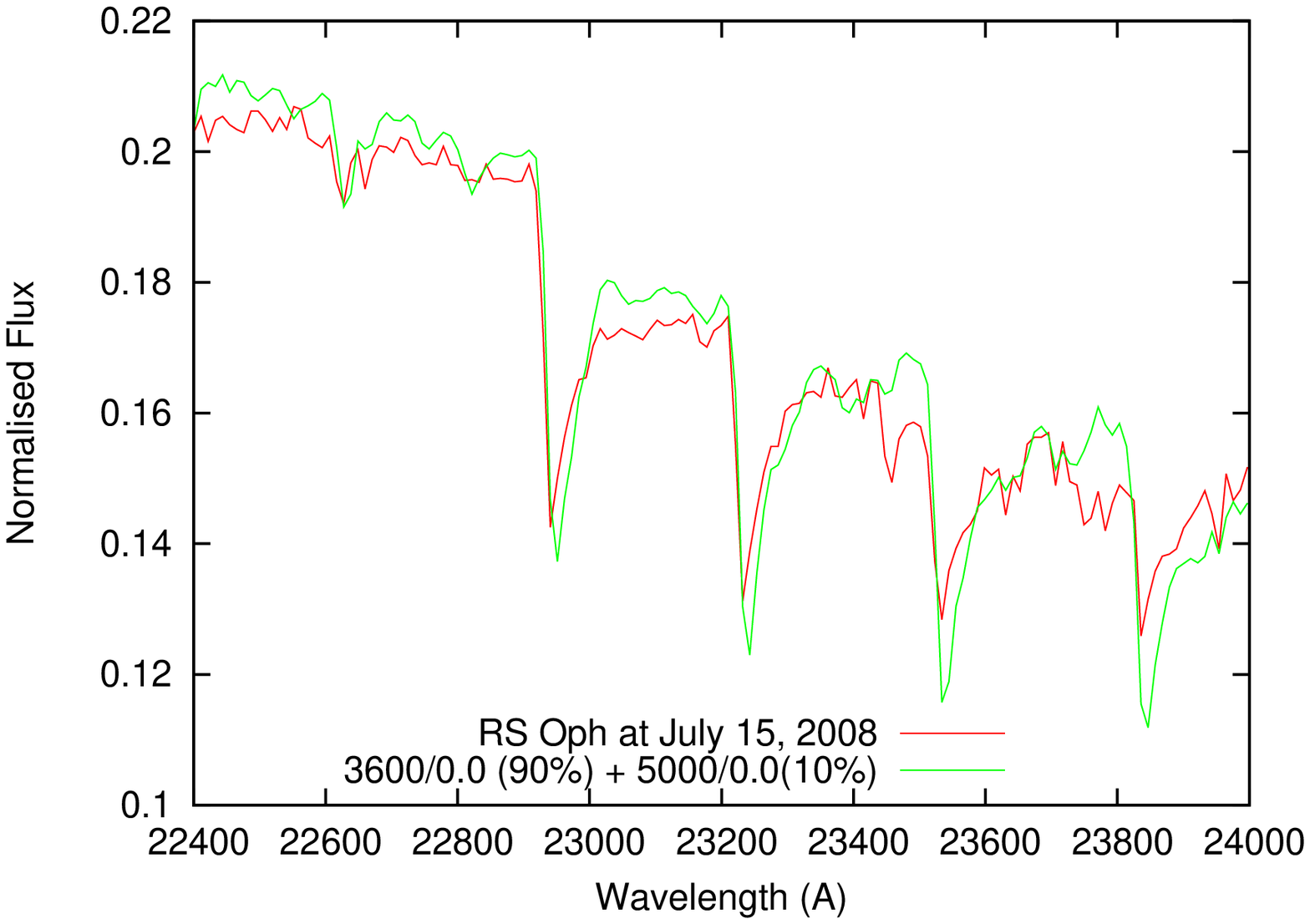}
      \caption{\label{_5000} Top: best fit to the AD-extracted IR spectrum
observed on July 15, 2008.
 with the two-component model: (90\% of 3600/0.0 and 10\% of 5000/0.0).
Middle: the fit of the optical spectrum from 2006 with the same model.
Bottom: The fit of two-temperature model spectrum to the CO bands observed on 2008
July 15. }
   \end{figure}

In the lower panel of Fig. \ref{_5000}, we show the 
fit to the observed CO bands. We see 
a qualitative agreement between theory and observations. In this case
we obtain the result with a model atmosphere having lower \Tef\ and \logg.
The results in Fig.~\ref{_5000} show rather a weak dependence 
on the adopted value \logg. We choose the fit with \logg = 0 because 
this value provides a better fit to some of the spectral detail in the
optical region. However, the resolution of
our spectra restricts our ability to draw any definitive conclusions: to
determine \logg\ we require echelle spectra.
In general the satisfactory  reproduction of the CO bands provides independent 
evidence that, for the 2008 dates for which we have data, \RS\ was in quiescent mode.
Any reduction in the  intensity of the CO
bands would imply changes in the temperature structure of the line 
forming region rather than the presence of chemical peculiarities.

Problems with reproducing the H$^-$ feature at 1.6\mic\ persist in our two-component model.
The weakness of the feature in \RS\
provides an argument in favour of the high temperature model. It is almost certain that
the actual structure of RG atmosphere is more complex than is described by
our two-temperature model. We will defer a study of these phenomena to forthcoming 
papers using more accurate optical and infrared data from the same epoch.

\section{Discussion and conclusions}

We have re-examined the IR spectrum of \RS\ obtained in 2008.
To  extract the contribution of the AD
from the observed SED we developed two models
of the AD: one, a single component model, with accretion 
rate $\sim3.9\times10^{-6}$\Mdot\, and a second, 
two component model, with a hot boundary layer and 
accretion rate $\sim2\times10^{-8}$ \Mdot. 
These models provide reasonable fits to the optical
spectral region for observation dates in 2007-2009. 

Our analysis shows that 
the SED
of the RG and the intensities of the emission lines vary on a short ($\ltsimeq1$~day) time-scale.
 We suggest that 
these changes are due to AD and/or hot boundary layer variability.

The extracted infrared spectra provide similar RG parameters 
for the two AD models.
After subtracting the
contribution of the AD, the IR spectrum observed on 2008 July 15 
yields a lower \Tef\ by comparison with our earlier work.
A fit to the observed spectrum in 2008 September confirms our finding of
a relatively low \Tef.  On the other hand,
a comparison of the optical data with the spectra of template stars 
suggests a much higher $\Tef =  4500\pm100$~K.

To explain the observed discrepancy we developed a more elaborate model,
in which there is a hot region on the surface of the comparatively
cool RG. The hot region
can most likely be explained by irradiation of the RG by the WD: the hotter 
part of atmosphere extends to the photospheric layers.

Below we summarise the main results of this paper:

\begin{enumerate}

\item The optical spectrum of \RS\ observed on 2008 June 25,
after extraction of contribution of AD, is similar
to those obtained in 2006, several months after the RN eruption.

\item the IR spectrum of \RS\ from 2008 July 15 consists of the
RG photosphere, together with a contribution from the AD.
To obtain the {\it photospheric} fluxes we used the same procedure to extract the AD flux
as for optical spectrum.

\item Our model is only valid if the physical state of the RG atmosphere
is determined mainly by its internal physical state. In this case, the visible 
variations of the system are determined by the changes in the AD.
The good agreement between the optical spectrum from 2006 and AD-extracted spectrum 
from 2008 provide some indirect confirmation of comparatively stable state of 
the photosphere of \RS. The observed short time variations are due to 
disc and/or hot boundary layer variability phenomena, not the processes in photosphere of the RG.


\item with our comparatively simple two-component model we can account 
for the main observed phenomena:

 \begin{enumerate}
\item the difference between the formal \Tef\ obtained from fits to 
optical and IR spectral energy distributions in the framework of the classical
modelling stellar spectra
\item the presence of strong TiO bands in the spectrum of the red 
giant with formally determined $\Tef=4500$~K for the classical 1D model atmosphere.
\end{enumerate}
\end{enumerate}

There are of course some remaining issues, yet to be investigated, with the
proposed two component model,
for example the dependence of the observed parameters on orbital phase.
By chance, our data were obtained near quadrature; further optical-IR spectra,
obtained at phase $\phi\simeq0.75$ when the irradiated hemisphere of the RG is
not visible, and $\phi\simeq0.25$ with the WD in front of the RG,
would be valuable to test our suggestion of an irradiated RG.
Such observations would be relatively easy to obtain. Also valuable, but
less straight-forward from a scheduling point of view, would be a long series of low resolution
observations at various phases, complemented by high resolution monitoring.

\section{Acknowledgments}

This work was supported by an International Joint Project Grant from the UK
Royal Society. YP's and BK's studies are partially supported by FP7 Project No: 
269193 ``Evolved 
stars: clues to the chemical evolution of galaxies". 
This work was co-funded under the Marie Curie Actions of the.
European Commission (FP7-COFUND).
This research has made use of the SIMBAD
database, operated at CDS, Strasbourg, France.
The United Kingdom Infrared Telescope was operated by the Joint Astronomy Centre
on behalf of the Science and Technology Facilities Council of the U.K.
 We thank the anonymous Referee for his/her thorough review and highly appreciate the comments and 
suggestions, which significantly contributed to improving the quality of the publication.

\end{document}